\documentclass[sigplan, nonacm, twocolumn,10pt]{acmart}
\renewcommand\footnotetextcopyrightpermission[1]{}
\author{Bernhard Kauer}
\affiliation{
    \institution{Eclipse Labs}
    \city{}
    \country{}
}

\author{Aleksandr Petrosyan}
\affiliation{
    \institution{Eclipse Labs}
    \city{}
    \country{}
}

\author{Benjamin Livshits}
\affiliation{
    \institution{Eclipse Labs, Imperial College}
    \country{}
}

\setcopyright{acmlicensed}
\copyrightyear{2025}
\acmYear{2025}
\acmDOI{XXXXXXX.XXXXXXX}

\usepackage{cleveref}
\usepackage{multirow}
\usepackage{bigstrut}
\usepackage{booktabs}
\usepackage{xspace}
\usepackage[T1]{fontenc}
\usepackage{ae,aecompl}
\usepackage{caption}
\usepackage{subcaption}
\usepackage{bigstrut}
\usepackage{amsthm}
\usepackage{titlesec}
\usepackage{amsmath}
\usepackage{algorithm2e}

\newcommand{\point}[1]{\par\smallskip\noindent\textbf{#1}. }

\newcommand{\etal}{\textit{et~al.}\xspace}
\newcommand{\tool}{\textsc{AlDBaran}\xspace}
\newcommand{\Circled}[1]{\raisebox{.5pt}{\textcircled{\raisebox{-.9pt} {#1}}}}
\titleformat{\subsubsection}[block]{\normalfont\bfseries}{\thesubsubsection}{0.6em}{}

\title{\tool: Towards Blazingly Fast State Commitments for Blockchains}

\newtheorem{observation}{Observation}

\begin{document}
\sloppy
\pagestyle{plain}
\begin{abstract}

The fundamental basis for maintaining integrity within contemporary blockchain systems is provided by authenticated databases. Our analysis indicates that a significant portion of the approaches applied in this domain fail to sufficiently meet the stringent requirements of systems processing transactions at rates of multi-million TPS (transactions per second). \tool signifies a substantial advancement in the realm of Authenticated Database Design. By eliminating disk I/O operations from the critical execution path, implementing prefetching strategies, and refining the update mechanism of the Merkle tree, we have engineered an authenticated data structure capable of handling state updates efficiently at a network throughput of 50 Gbps. This throughput capacity significantly surpasses any empirically documented blockchain throughput, thereby guaranteeing the ability of even the most rapid and high-throughput blockchains to generate state commitments effectively.

Specifically, \tool provides support for historical state proofs, which facilitates a wide array of novel applications. For instance, the deployment of \tool could enable blockchains that do not currently support state commitments to offer functionalities for light clients and/or implement rollups. Additional applications might encompass the aggregation of Time-Weighted Average Prices at the forefront of the blockchain with any desired level of temporal precision.

When benchmarked against alternative authenticated data structure projects, \tool exhibits superior performance and simplicity. In particular, \tool achieves speeds that are at least ten times faster than QMDB, handling approximately 48 million updates per second using an identical machine configuration. Furthermore, \tool relies on only three external packages, compared to QMDB's requirement for 21 external dependencies. This characteristic renders \tool an attractive solution for resource-limited environments, as its historical data capabilities can be modularly isolated (and deactivated), which further enhances performance. Demonstrations have been conducted on consumer-level portable hardware, achieving approximately 8 million updates per second in a purely in-memory setting and 5 million updates per second with the ability to perform snapshots at sub-second intervals, thereby illustrating compelling and cost-effective scalability.
\end{abstract}

\maketitle

\section{Introduction}

Modern blockchains serve as the backbone of many financial, physical infrastructure, and entertainment services.  As of August 2025 total value locked in decentralized finance (DeFi) is approximately \$141 billion\footnote{\url{https://defillama.com/}}.  Render, a famous decentralized physical infrastructure network touts near unlimited cloud GPU rendering power.  It is often the case that the cost of these services is dominated by the technical characteristics of the backing blockchain, with performance, particularly of authenticated state management, central among them.

When analyzing the performance of programmable blockchains, it is often the case that transaction (smart contract) execution is the bottleneck.  Further breakdown reveals that among the blockchains that feature an authenticated data store (ADS), as much as 81\% of the execution time can be spent performing storage operations~\cite{lvmt,rise}.  Poor performance of ADS is also the key reason behind their notable absence in Solana~\cite{solana}, one of the largest high throughput, parallel layer-1 blockchains.

It is worth revisiting the role of an authenticated data store in a typical blockchain.  Ethereum is a permissionless layer 1 (L1) blockchain  featuring two distinct types of node: full and light clients.  Full nodes perform the data processing and participate in consensus.  Light clients typically only retain the block headers and  a subset of the ledger state.  Omitting transactions obviates the need for light nodes to re-execute them, but requires some other mechanism for state verification.

In Ethereum this is accomplished by the main nodes computing a \emph{state commitment}, which is typically the root of a Merkle-Patricia trie corresponding to the blockchain state and is included in the block header.  A full node is thus able to generate proofs of state to the light nodes.  The Ethereum ledger state is thus \emph{authenticated}.  Unfortunately, Merkle-Patricia tries require \(O(\log n)\) computations per single state modification, for \(n\) indexed state entries.  For instance, assuming the ledger consists of \(1,024\) entries, a single transfer would require \(19\) additional operations\footnotemark.
\footnotetext{We decrement the source and increment the target account balances;  {\(\log 1024 = 10\)} and the root node can be updated lazily.}

\point{Challenges of high-speed blockchains}
% Explain why the expectations in terms of latency are different; explain snapshots and snapshot frequency. Set the stage for how fast things must be. 
The authenticated database's primary goal is to compute the state commitment for every block.  Clearly, the amount of work that needs to be performed depends on the number of state changes, their ordering, and the latency budget dictated by the block production cadence.  Obviously, if there are no state changes, the previous root can be re-used.  As another example, if two adjacent leaf nodes are updated, the amount of work needed to re-compute the root is only one extra operation, eliding the need for an expensive \(O(\log n)\) traversal.  Similarly, if the same key is updated repeatedly within a block, deferring recomputation can avoid redundant traversals.

Solana is a high performance blockchain that executes multiple transactions in parallel, producing multiple blocks per second.  Contrast this against Ethereum that produces one block every~12 seconds or more which are filled with transactions that can only be executed sequentially, and in a virtual machine that offers less performance than the modified eBPF that is used in Solana.  Because the ADS is operating under a much tighter latency budget, and because there are multiple concurrent updates of several non-adjacent leaf nodes at any moment in time, the ADS faces a tighter latency budget and higher concurrency pressure even when state sizes are comparable.

To support a theoretical blockchain with one million transactions per second, a 400 millisecond block time, and roughly the same workload as the Solana main network: 3 account updates per transaction, the authenticated database must be able to handle an average of three million updates per second.  The tree must sustain~3 million updates per second, with continuous root-path recomputation at that rate.  In an naive implementation, this work competes for the system resources needed to operate the node, \textit{e.g.} RAM and CPU cores. 

Additionally, an authenticated database in a normal setting needs to periodically persist the blockchain state.  This state is called a \emph{snapshot} and the process is called \emph{snapshotting}.  Snapshot cadence can be decoupled from block time.  It is largely dictated by the rate of state change and by the need to produce inclusion/exclusion proofs for recent blocks.  Some deployments omit frequent snapshots, but their utility for proofs and recovery often outweighs the overhead.  As such it is often the case that the snapshots are taken every few blocks.  In principle, the cadence can be dynamically adjusted. 

\point{Using fast state commitments}
%Talk about bridging, light nodes, trusted RPC settings, etc.
The applications of fast state commitments are numerous.

In the context of bridging, \textit{i.e.} connecting multiple blockchains for cross-chain operations, such as token exchanges, the fast state commitments can significantly reduce operation latency.  Cross-chain communication often connects chains with different consensus mechanisms, block production cadences, and finality rules.  Per-block commitments reduce bridging latency; more frequent commitments than the block cadence can enable optimistic paths when cadences are mismatched (\textit{e.g.}, Ethereum and Solana).

For trusted RPC nodes, efficient commitments improve responsiveness.  Fast commitments provide early confirmation of a transaction’s effects and serve as proofs of the current state; clients can use this information to formulate transactions more accurately.

Light clients are clients that do not instantiate the full blockchain state, but instead operate on a small subset thereof.  For light clients to be useful, they must be able to reason about sufficiently up-to-date information from the blockchain.  For Ethereum-like chains, because of the relatively small amount of changes, the long block production cadence and built-in support for state commitments, light clients are commonplace.  This is not true of Solana.  As we have mentioned earlier, Solana is an adverse environment for authenticated databases by virtue of a low latency budget and parallel execution.  Solana does not employ a production ADS in its runtime; as a result, native light-client support is limited and production-grade light clients are not widely deployed. Consequently there are no light clients for Solana.  This is especially problematic, because the system requirements for operating a full node are already much higher than for \textit{e.g.} Ethereum.  A fast authenticated database would unlock light clients for high performance blockchains. 

\subsection{Contributions}
This paper makes the following contributions.
\begin{enumerate}
\item Our main contribution is a novel architecture that completely removes persistent I/O from the hot execution path, thus minimizing latency.
\item We propose a concurrent, depth-limited, asynchronous update approach for the in-memory (sparse) Merkle tree, enabling best-in-class throughput for state-root generation.
\item We develop an allocation strategy enabling efficient prefetching of Merkle tree nodes into the CPU caches, for  efficient CPU-based state root generation.
\item We further show how to generate fast post-hoc inclusion and exclusion proofs.
\item Finally, we provide a reference implementation as a Rust library with minimal dependencies.
\end{enumerate}
These contributions are demonstrated in benchmarks using AWS machines and show a practical throughput suitable for high performance applications:~48 million state updates per second for a representative workload.  Assuming three state updates per transaction, this roughly translates to~16 million transactions per second. 

\subsection{Organization}
The rest of the paper is organized as follows.
Section~\ref{sec:background} provides some information on related authenticated database systems that are used in the blockchain context.
Section~\ref{sec:overview} presents an overview of \tool{} and its technical foundations.  The two subsequent sections:(\ref{sec:data-organization} and \ref{sec:optimizations}) detail the data structures as well as the specific optimizations employed in \tool{}.
Section~\ref{sec:data-organization} talks about data organization and storage.
Section~\ref{sec:optimizations} discusses various optimizations.
Section~\ref{sec:proofs} presents \tool{}'s proof system.
In Section~\ref{sec:eval} we provide an experimental evaluation.
Section~\ref{sec:discussion} provides discussion of some of the relevant issues highlighted by our experiments.
Section~\ref{sec:related} presents related work.
Finally, Section~\ref{sec:conclusions} concludes.

% Local Variables:
% jinx-languages: "en_US"
% TeX-master: "main"
% End:

% TODO: Fraud proof creation

\begin{figure*}[tbp]
    \centering\footnotesize
    \begin{tabular}{|p{3cm}|p{2.5cm}|p{2.5cm}|p{2.5cm}|p{2.5cm}|p{2.5cm}|}
    \hline\bigstrut[t]
    \textbf{Dimension} & \textbf{Firewood (Avalanche)} & \textbf{MerkleDB (Avalanche)} & \textbf{NOMT (Nearly-Optimal Merklization)} & \textbf{LVMT (Multi-Layer Versioned Trie)} & \textbf{QMDB (Quick Merkle Database)} \bigstrut[t]\\
    \hline\hline
    \textbf{Architecture \& Data Model} & Integrated on-disk Merkle Trie: stores trie nodes directly with a $B^+$-tree-like layout & Layered Merkle Radix Trie on top of RocksDB & Binary Merkle Trie + separate flat KV store for values & Vector-Commitment Merkle Tree + Authenticated Multipoint Tree (AMT) overlay & Append-only “twigs” (fixed-size subtrees) unifying KV + Merkle storage \bigstrut[t]\\
    \hline
    \textbf{Storage Integration \& I/O} & Compaction-less, purely sequential writes; eliminates LSM write-amplification & RocksDB LSM with leveled compaction; each trie op → O(log N) I/Os & Flash-native layout; predictable 1–2 I/Os per lookup/update & Constant-time root update via vector commitments; still multi-I/O for proofs & O(1) writes (batched by 2,048 entries), 1 SSD read per lookup \\
    \hline
    \textbf{Merkle Computation} & On-disk trie writes and in-place hashing; WAL for atomic commits & Merkle nodes persisted as KV pairs; proof generation incurs disk reads & Standard Merkle updates with multiproof support & Merkle root updates in O(1) via vector commitments; proofs via AMT lookups & Fully in-memory Merkleization of twigs—no disk I/O for hashing \\
    \hline
    \textbf{Versioning \& Historical} & Tracks only recent revisions (configurable N); older states pruned in-place & Copy-on-write Views + RocksDB snapshots; GC of old nodes & Single-state design; no built-in historical queries & Native support for exclusion/latest-value proofs via AMT, but not full history & Full historical proofs (inclusion/exclusion at any block) via OldId pointers \\
    \hline
    \textbf{Peak Throughput} & Targets high TPS with very low latency in alpha & Handles Avalanche’s high TPS; bounded by RocksDB compaction spikes & About 43k updates/sec per thread in PoC; SSD-bound benchmarks & Benchmarks show about 30\% uplift vs MPT in average case; constant-time roots & Up to 2.28M updates/sec on NVMe RAID; 1M TPS in token tests \\
    \hline
    \textbf{Maturity \& Production} & Alpha/dev preview, planned mainnet integration, restrictive license & Production in AvalancheGo since late 2023; edited BSD-licensed & Proof-of-Concept; early Rust prototype (MIT/Apache) & Research prototype (OSDI ’23); full integration pending & Research prototype; lab benchmarks only to date, MIT/Apache license \\
    \hline
    \textbf{Reusability} & General-purpose Rust crate; could serve any Merkle-trie chain & Embedded in Go Avalanche nodes; reusable within Go ecosystem & Pluggable in Substrate/rollups; unopinionated on key format & Applicable to any blockchain needing fast proofs; vector commitments portable & Generic KV+ADS library in C++; twig concept portable to other contexts \\
    \hline
    \end{tabular}
    \caption{Comparison to closely related projects. More academic work is described in Section~\ref{sec:related}.}
    \label{fig:comparison-table}
\end{figure*}
\section{Background}%
\label{sec:background}

%\subsection{The Challenge: Beyond Layered Architectures}
Blockchain state management traditionally combines a proof layer (an ADS) with a storage layer (a key-value store).  The classic implementation, seen in Ethereum with its Merkle Patricia Trie (MPT) on top of LevelDB or RocksDB, suffers from significant I/O overhead.  This compaction-based model introduces significant write amplification, often \(5-10\times\) more physical writes per logical update, and read amplification, since each operation must touch multiple Log Structured Merge Tree~(LSM) levels and trie nodes.  Each state update can cause a cascade of disk operations, leading to an \(O\left[{(\log N)}^2\right]\) I/O cost that makes the execution layer I/O-bound.

Emerging blockchain architectures demand ADS engines that rethink the separation between ``proof'' and ``storage''. The ideal design minimizes write amplification by pruning outdated data in place or by batching updates.  It targets \(O(1)\) or low-constant I/O complexity to avoid I/O-bound storage layers,  keeps memory footprint reasonable, trading off in-memory hashing against on-disk commitments where appropriate, as well as extends proof capabilities beyond membership to range, change, and historical queries.  Finally, production-readiness, based on code maturing, audit history, and ecosystem integration, determines how readily an existing design can be adopted in live networks.

Addressing these challenges, a new generation of integrated ADS engines has emerged.  Each engine offers a unique approach to overcoming these bottlenecks, as we outline below and go into more depth in Section~\ref{sec:related}. A table comparing these solutions is provided in Figure~\ref{fig:comparison-table}.

\subsection{Avalanche MerkleDB}
Of the modern authenticated databases, Avalanche's {MerkleDB} is the most ``traditional'', as it retains the classic design of layering a Merkle trie on top of a generic database like RocksDB, representing a highly optimized version of the layered approach.  It has served as the backbone of the Avalanche network, proving itself capable of handling high-throughput workloads and sub-second finality.

As we mentioned before, MerkleDB offers a more conservative evolution, layering a Merkle Radix Trie atop RocksDB.  It leverages copy-on-write ``Views'' for provisional state mutations and batches trie commits to reduce round-trip I/Os.  Compression algorithms like Snappy and Zstandard~(zstd) mitigate storage growth, while RocksDB's mature compaction tuning enables sustained sub-second finality.

MerkleDB reliably supports several thousand TPS in production and was audited by OpenZeppelin in March~2023\footnotemark, signaling high confidence in its stability.  However, because it preserves the underlying LSM design, MerkleDB inherits the non-trivial write amplification factor and background compaction overhead, making it less I/O-efficient than fully integrated engines.
\footnotetext{\url{https://blog.openzeppelin.com/merkledb-audit}}

\point{Architecture and Data Model}
It is written in Go and is tightly integrated into the AvalancheGo client.  Within its model, each trie node is stored as a distinct entry in the database, keyed by the node's hash.  To improve efficiency, MerkleDB separates value-bearing leaf nodes from intermediate branch nodes using distinct key prefixes within RocksDB, which simplifies iterating over the actual state data.

\point{Proof Scheme}
MerkleDB is built on a standard Radix-16 Merkle Trie (akin to Ethereum's Patricia Trie).  Keys are split into 4-bit nibbles, so each lookup or update walks at most \[\lceil \text{key\_length}/4 \rceil\] levels, and proofs consist of a single sibling-hash per level.  This delivers \(O(d)\)-sized (where \(d\) is trie depth) inclusion and non-inclusion proofs, minimal proof-overhead, and constant-time hash verifications, ensuring both compactness and cryptographic soundness.

A key architectural feature of MerkleDB is its use of copy-on-write ``Views.'' A View is a lightweight, in-memory snapshot of the trie where modifications can be applied speculatively.  This is crucial for block execution, as it allows the system to process transactions and even stage changes for pending blocks deferring  mutating the on-disk state.  Only when a block is finalized are the changes from its corresponding View committed and merged into the base trie in a single, batched operation.

\point{Performance and Inherited Trade-offs}
The View mechanism allows for a degree of optimistic concurrency, but commits are serialized through a \texttt{commitLock} to ensure atomicity, effectively rendering the storage layer into a single-writer high-contention system during block application.

While MerkleDB is battle-tested and production-ready, (having been audited by OpenZeppelin in March 2023 as we mentioned earlier), it cannot escape the fundamental trade-offs of its layered design.
In summary, MerkleDB's use of a radix-16 Merkle Trie guarantees compact, depth-bounded proofs and \(O(d)\) proof generation, offering stronger verifiability than schemes with larger or variable-sized proof structures.  Its straightforward Merkle design delivers both efficiency and robust cryptographic guarantees.

MerkleDB represents the apex of the layered model.  Its inefficiencies are intrinsic to the model itself and to achieve better throughput, one must consider a different architecture altogether.  This is what inspired the creation of the successor ADS called Firewood.

\subsection{Avalanche Firewood}
Developed by Ava Labs as the next-generation storage engine for the Avalanche ecosystem, \textbf{Firewood} substantially departs from the layered design of MerkleDB.  It is a purpose-built, integrated authenticated database that reduces the overhead of a separate key–value abstraction by storing trie nodes natively.

Firewood reconceptualizes the state trie as the primary on-disk index, organizing nodes in a \(B^{+}\)-tree-like, compaction-free structure.  Obsolete trie revisions are removed in situ via a ``Future-Delete Log'' eliminating the necessity for background compactions or layered key-value systems.  By employing the trie as the main data structure, Firewood significantly decreases write amplification and reduces random I/O.  Although integration with proof systems is planned, Firewood currently excels in aligning trie I/O directly with SSD semantics, avoiding the inefficiencies inherent in LSM compactions.

\point{Eliminating the Layers}
Firewood's core innovation is simple but profound: the on-disk index is the Merkle trie itself.  Instead of flattening the trie structure into key-value pairs, Firewood stores trie nodes natively on disk in a B$^{+}$-tree-like layout.  This eliminates the expensive serialization and abstraction layers, aligning the physical storage directly with the logical data structure.  It is a compaction-less engine written in Rust, designed from the ground up to minimize I/O and maximize throughput.

\point{Architecture and Storage Management}
Firewood performs in-place updates on the active state and prunes outdated data on the fly, avoiding the write amplification that plagues LSM-based systems like RocksDB.  When a block is committed, new trie nodes are written to disk, and any nodes that become obsolete are recorded in a Future-Delete Log (FDL).  When a state version (or ``revision'') expires and falls outside the configured retention window, Firewood processes the FDL to reclaim the space occupied by those stale nodes.  This clever mechanism avoids the need for a separate, resource-intensive pruning or compaction process.

The design is highly optimized for modern SSDs.  By co-locating nodes that are frequently accessed together (e.g., nodes along a path in the trie), Firewood improves cache locality and reduces disk seeks, a technique known as location-aware storage.

\point{Performance and Goals}
Firewood is engineered to meet the high-throughput requirements of Avalanche and its subnets by storing the trie itself in a compaction-less B$^{+}$-tree layout.  Crash recovery is managed via a standard write-ahead log (WAL), and by eschewing an LSM-tree design, Firewood avoids the write stalls and throughput degradation typically caused by compactions.  There were some performance comparisons against Geth, but we could not find any TPS numbers from Ava labs on this.

\point{State History, Licensing, and Status}
Firewood maintains a partial state history through its ``revision'' system, retaining a configurable number of recent state versions while garbage-collecting older ones.  This provides fast access to recent states for rollbacks or syncing without requiring unbounded disk growth.

Currently in an alpha developer preview, Firewood is still under active development.  A significant consideration for the broader community is its licensing.  The code is available under the ``Ava Labs Ecosystem License v1.1'' which restricts its use to the Avalanche public networks.  This source-available but restrictive model is designed to give the Avalanche ecosystem a competitive advantage.

In summary, Firewood is a specialized state engine that trades generality and permissive licensing for a massive leap in single-node performance, showcasing the power of a truly integrated ADS design.

\subsection{Nearly-Optimal Merklization (NOMT)}
Developed as a Rust-based research prototype, \textbf{NOMT} pairs a binary Merkle Trie with a flat, page-aligned key-value store to reduce random I/O on SSDs.  By aligning nodes to flash page boundaries and avoiding pointer-chasing within the trie, NOMT achieves approximately~43,000 state updates per~second per~thread, fully saturating modern NVMe throughput at multi-gigabyte scales.  While NOMT cannot eliminate the fundamental~$O(\log N)$ trie height, it reduces constant factors dramatically, yielding an order-of-magnitude improvement over naive MPT implementations.

Nearly-Optimal Merklization (NOMT) is less a specific product and more a design philosophy that has deeply influenced the development of modern authenticated databases.  Championed by researchers like Preston Evans and Polkadot's Robert Habermeier, NOMT's central thesis is that performance gains come from aligning the data structure's physical layout with the characteristics of the underlying hardware, particularly SSDs.

\point{The Core Idea}
NOMT proposes two fundamental shifts.  First, it moves from a high-arity radix trie (like MerkleDB's Radix-16) to a simple binary Merkle tree (arity~2).  While binary trees are deeper, their proofs are smaller, requiring only one sibling hash per level instead of up to 15.

Second, and more importantly, NOMT decouples the logical trie structure from its physical on-disk layout.  It achieves this by packing multiple binary trie nodes into a single, fixed-size disk page (e.g.,~4~KB), which is the native block size for SSDs.  A single~4~KB page can hold a complete binary sub-trie of depth~6, containing~64 nodes.

This was a significant step.  A lookup or update in a traditional trie might require dozens of dependent, random disk reads to traverse the tree from root to leaf.  With NOMT, that same operation might only require fetching one or two~4~KB pages from disk, an order-of-magnitude reduction in random I/O.

\point{Performance and Data Model}
Because the location of the required pages can be calculated directly from the key's bits, the system can issue all necessary page reads in parallel, fully exploiting the capabilities of modern NVMe SSDs.  Early benchmarks of a NOMT prototype demonstrated roughly~50,000 trie updates per second, a significant speedup over naive implementations.

To handle the variable-length keys common in blockchains, the Polkadot adaptation of NOMT uses a padding scheme to create uniform-length keys and introduces extension nodes to efficiently represent long, shared key prefixes without breaking the page-aligned layout.

\point{Status and Influence}
NOMT is primarily a set of research and development concepts rather than a deployed product.  However, its ideas are highly influential and have been cited by teams working on other high-performance databases, including MegaETH.  It represents a fundamental rethinking of how to persist authenticated data structures, prioritizing I/O efficiency above all else.  By trading a small amount of space inefficiency (not all pages will be full) for a massive reduction in disk seeks, NOMT provides a blueprint for saturating modern storage hardware.

\subsection{Layered Versioned Multipoint Trie (LVMT)}
{LVMT} couples an append-only Merkle tree with an Authenticated Multi-point Trie underpinned by algebraic vector commitments.  Instead of hashing every branch upon each update, LVMT stores compact commitment data in the trie and defers the bulk of cryptographic work to vector-commitment operations that run in amortized~\(O(1)\) time.  In Ethereum-like benchmark scenarios, LVMT delivers up to~$6\times$ faster read/write performance and~\(2.7\times\) higher overall transaction throughput compared to conventional MPT on LSM storage.  Its primary trade-off is the complexity of the underlying commitment scheme, which demands specialized libraries and, in some variants, a trusted setup.

While NOMT tackles the performance problem at the hardware layout level, the Layered Versioned Multi-point Trie (LVMT) approaches it from a cryptographic and algebraic angle.  Presented in a~2023 OSDI paper, LVMT is a novel design that leverages more advanced cryptographic primitives to change the asymptotic complexity of state updates.

\begin{figure*}[tb]
  \centering
  \includegraphics[width=\linewidth]{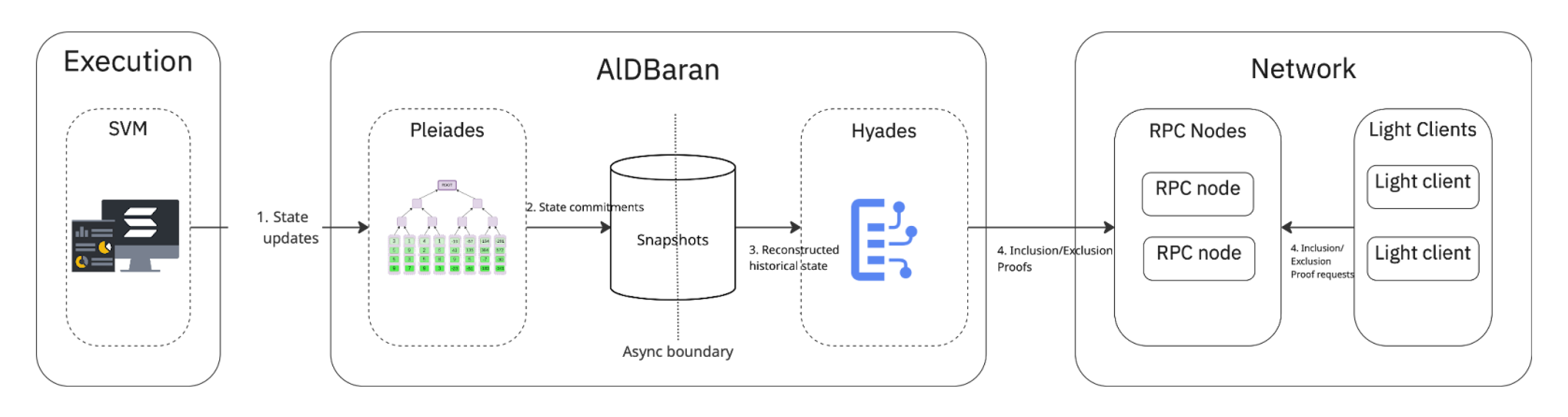}
  \caption{Architecture of \tool.}%
  \label{fig:arch}
\end{figure*}

\point{Vector Commitments}
LVMT's architecture is built on a different type of cryptographic accumulator called a vector commitment.  Specifically, it uses an Authenticated Multi-point Evaluation Tree (AMT) at its base layer.  An AMT allows for committing to a large array of values in a way that allows for extremely efficient updates.  Instead of re-hashing \(O(\log N)\) nodes up to the root for every change, an AMT can often update its commitment in constant time using techniques like polynomial commitments.

\point{Architecture}
LVMT is not a single monolithic structure.  It intelligently layers an append-only Merkle tree on top of these powerful AMT structures.  The bottom layer consists of multiple AMTs, each managing a segment of the total key space.  The upper layers of the trie do not store raw data or even full hashes of the layers below.  Instead, they store much more compact data, such as simple version numbers.

When a value in a bottom-layer AMT is updated, the expensive cryptographic work is contained within that AMT.  The change propagates up to the higher layers merely as an incremented version number, not a cascade of new hashes.  This clever design avoids performing expensive elliptic curve operations or other heavy cryptography across the entire tree for every small update.

\point{Asymptotic Breakthrough and Performance}
This layered, versioned design leads to a profound performance breakthrough: amortized \(O(1)\) root commitments.  The cost of updating the state commitment becomes nearly constant, regardless of the total state size, allowing LVMT to sidestep the $O(\log N)$ hashing bottleneck of traditional Merkle tries.

The practical results are impressive.  In experiments running on Ethereum-like workloads, LVMT delivered up to~\(6\times\) faster read/write operations and boosted overall transaction throughput by~\(2.7\times\) compared to a conventional MPT on LSM storage.  The design is also inherently versioned, making it naturally suited for historical state queries.  A Rust implementation of the underlying AMT is being developed in the Conflux ecosystem, suggesting a path toward production use.

LVMT represents the frontier of authenticated data structures, showcasing how a shift in the underlying cryptographic tools can fundamentally alter the performance landscape.

\subsection{Quick Merkle Database (QMDB)}
{QMDB} unifies key-value storage and Merkleization in fixed-size, append-only ``twig'' structures, each batching 2,048 entries.  By grouping updates into twigs, it achieves~\(O(1)\) SSD I/O per state change and a single SSD read for lookups, while maintaining in-memory Merkle hashing at just~2.3 bytes of RAM per entry.  Benchmarks show sustained throughput of up to~2.28 million updates per second, supporting upwards of one million TPS in ideal conditions, and practical scaling to over~15 billion entries.  QMDB also provides built-in historical proofs, enabling past-state queries for auditing and analytics.

If the previous designs represent powerful new takes on specific components of the state problem, the Quick Merkle Database~(QMDB) represents their synthesis, a hyper-optimized engine that unifies the key-value store, Merkle tree, and versioning system into a single, cohesive architecture.  Developed by LayerZero Labs and open-sourced, QMDB sets a new standard for performance in authenticated databases.

\point{Architecture}
QMDB completely collapses the traditional two-layer stack into a single, integrated data structure.  There is no separate KV store; keys, values, and all Merkle metadata are persisted together in one unified, append-only log structure.  This eliminates redundant storage and the communication overhead between layers, which is a major source of its~\({6-8\times}\) performance gain over systems like RocksDB+MPT.

The central innovation in QMDB's design is the ``twig'' A twig is a fixed-size Merkle sub-tree that holds a batch of 2,048 leaf entries.  All state updates are first buffered into a twig in memory.  The Merkle root of this small twig is updated on the fly, entirely in RAM.  Only when a twig is full is its content flushed to disk as a single, contiguous, append-only block.

\point{Performance and Efficiency}
This design has some significant performance implications:
\begin{itemize}
    \item \textbf{Optimal I/O:} State updates require only \(O(1)\) disk I/Os (a single sequential write per flushed twig), and lookups require at most a single SSD read.  This is the theoretical minimum I/O complexity.
    \item \textbf{In-Memory Merkleization:} All Merkle hashing is done in memory, meaning there is zero disk I/O overhead for calculating state roots during block execution.
    \item \textbf{Minimal memory footprint:} This is achieved with an incredibly low memory footprint of just \(\sim2.3\) bytes of DRAM per entry for the Merkle state, thanks to a highly compressed representation of the twigs' metadata in memory.
    \item \textbf{Extreme throughput and scalability:} In benchmarks, QMDB~\cite{qmdb} achieved up to~2.28 million updates per second and sustained~1M TPS on a token transfer workload.  It has been tested with over~15 billion entries, ten times the size of Ethereum's current state, on a single server.
\end{itemize}

\point{Historical Proofs}
QMDB is a fully versioned database from the ground up.  Every entry contains pointers to its previous version, forming a linked list of changes over time.  This allows QMDB to offer a powerful and unique feature, historical proofs.  A user can query the database for the value of a key at any past block height and receive a valid Merkle proof against that historical state root, all while querying the most recent version of the database.  This capability unlocks new application possibilities, such as on-chain verification of past events, without relying on trusted archival nodes.

QMDB stands as the current state-of-the-art, demonstrating what is possible when the data structure, storage layout, and cryptographic process are co-designed for maximum efficiency on modern hardware.

Taken together, these systems trace a path from layered designs (MerkleDB), to integrated engines (Firewood), to hardware-aligned layouts (NOMT) and cryptographic abstractions (LVMT), and finally to unified designs (QMDB). A common thread remains: state updates and persistence for historical queries are tightly coupled. This motivates a design that fully separates the in-memory hot path from asynchronous persistence and proof generation, which we pursue in section~\ref{sec:overview}.

% \subsection{AlDBaran vs.  QMDB}
% QMDB claims~2.28M updates per sec; when we re-ran their v0.2.0 code on the same AWS setup, we only saw~1.28 million updates per sec (and~0.803 million updates per sec on the June~26 2025 release).  By contrast, AlDBaran's Pleiades hot-path consistently outperforms both versions by over~$2\times$, underscoring its superior scalability and reduced contention.

% Even with a more frequent commit period, AlDBaran is over an order of magnitude faster than the highest throughput we could measure from the public QMDB code.  This direct comparison solidified our conclusion: to meet the demands of a~1M TPS rollup, a new architecture was not just an option, but a necessity.

% Local Variables:
% jinx-languages: "en_US"
% End:

\section{Overview of \tool}%
\label{sec:overview}

As illustrated in Figure~\ref{fig:arch}, at its core, \tool{} is designed to be located between the blockchain execution layer and the network-facing RPC/light-client layer, breaking down its state management into two purpose-built subsystems, \emph{Pleiades} for nanosecond scale in-RAM updates and Hyades for inclusion and exclusion proof generation. These two components are bridged through the use of \emph{snapshots}.
When the execution engine completes a block, three streams proceed in parallel:
\begin{enumerate}
\item \textbf{State updates:} (Execution engine $\rightarrow$ Pleiades)
As each $\langle$key, new-value$\rangle$ update arrives, Pleiades immediately applies it to its sparse, twig-sharded Merkle tree entirely in DRAM.  Updates and root-hash computation run completely asynchronously to execution: once all pending updates are applied, Pleiades emits a single~32-byte root hash.

\item \textbf{State commitment:} (Pleiades $\rightarrow$  Snapshots)
The state updates are then processed at the snapshot storage layer, where we \emph{compact} the data for later usage in Hyades. This process is asynchronous, meaning that it can be computationally expensive, giving us some flexibility to schedule it as eager or lazy computation.

\item \textbf{Reconstructed historical state:} (Snapshots $\rightarrow$ Hyades)
Hyades ingests transactions into an append-only proof log. Here, indexing by block height prepares the historical state for consumption by RPC nodes.

\item \textbf{Inclusion proofs:} (Hyades $\rightarrow$ RPC nodes)
The RPC nodes obtain snapshots from and/or run Hyades. They generate inclusion/exclusion proofs to light clients. Snapshots can be distributed using one of many available multicast propagation mechanisms.   
\end{enumerate}
This split architecture prevents hot-path operations in Pleiades (hashing, in-RAM updates) from being blocked by disk I/O or global locks, while archival and proof-generation processing runs asynchronously in Hyades, guaranteeing tens of millions of state updates per second.

The code representation of this design is a minimal dependency re-usable Rust library (\textit{i.e.} crate) that is suitable for \texttt{no\_std} environments such as inside kernels, trusted execution environments, as well as zero-knowledge prover systems such as RISC-0. Our implementation has been tested on \texttt{x86\_64}, \texttt{aarch64}, \texttt{RISC-V}, as well as their~32-bit counterparts: \texttt{i686} and \texttt{riscv32im}.

\subsection{Design Principles}
\subsubsection{Pleiades}
To hit multi-million state updates per second on the hot path, Pleiades follows the following design principles:
\point{Lock-free Concurrent Updates} the key space is split across cores, thereby eliminating locking from the critical path. Additional SIMD-friendly update grouping allows us to hash 16 tree nodes in parallel.

\point{Deterministic Breadth-first Layout with Prefetch Hints}  This is to keep the next node in the L2 cache by the time it is to be processed. 

\point{Lazy Root Computation} Per-update hashing stops at a per-thread subtree root; the top of the tree is updated once per block.

\point{Lazy Disk Writes} Every byte lives in DRAM until the block is committed.  Because 

\point{Asynchronous snapshots} the snapshot frequency is independent of block time.  Snapshots can be produced multiple times per block, providing intra-block state proofs. 

\subsubsection{Hyades}
Hyades in turn can use the snapshots at its own pace, creating inclusion and exclusion proofs based on the submitted external queries. The raw data for the state updates can be similarly stored separately, in a desegregated fashion, from the snapshot data. This would allow us to completely eliminate I/O operations from the critical path, resulting in Hyades being able to serve inclusion and exclusion proofs within milliseconds, without blocking.
Because of desegregation, Hyades can afford to be disk-bound. As with other components, it could run on completely separate machines in a data center context. We could also do more computationally intensive work in Hyades: for example, Hyades could go beyond inclusion and exclusion proofs and provide range proofs as well.

\begin{figure}[tb]
  \begin{subfigure}{\linewidth}
    \centering
    \includegraphics[width=.75\linewidth]{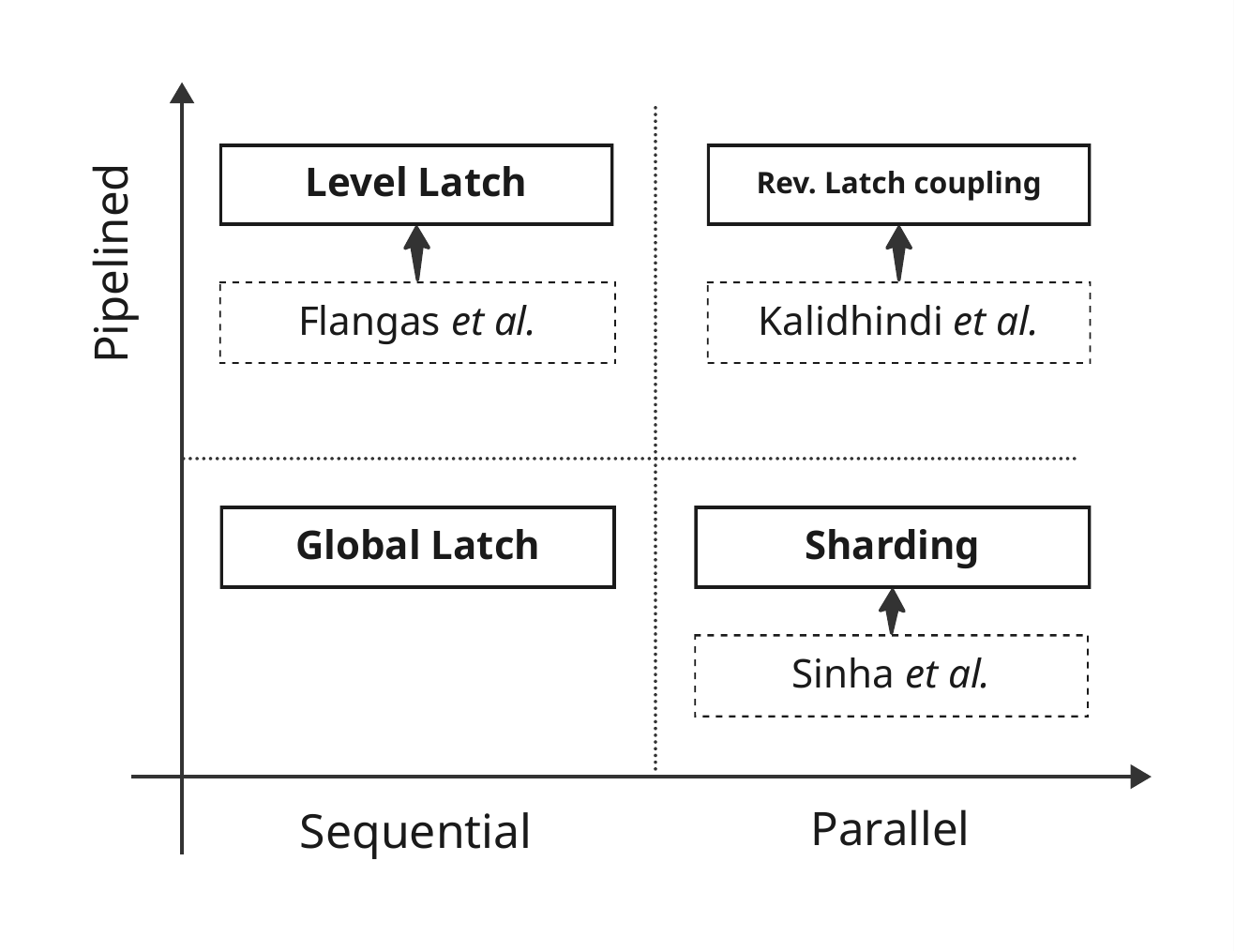}
    \caption{An overview of the design space for concurrent Merkle tree updates.  The discussion is in terms of pipelined execution, (the \(y\)-axis) and parallel execution (\(x\)).}
    % TODO: Add citations to these papers
  \end{subfigure}

  \begin{subfigure}{\linewidth}
    \centering
    \includegraphics[width=1.05\linewidth]{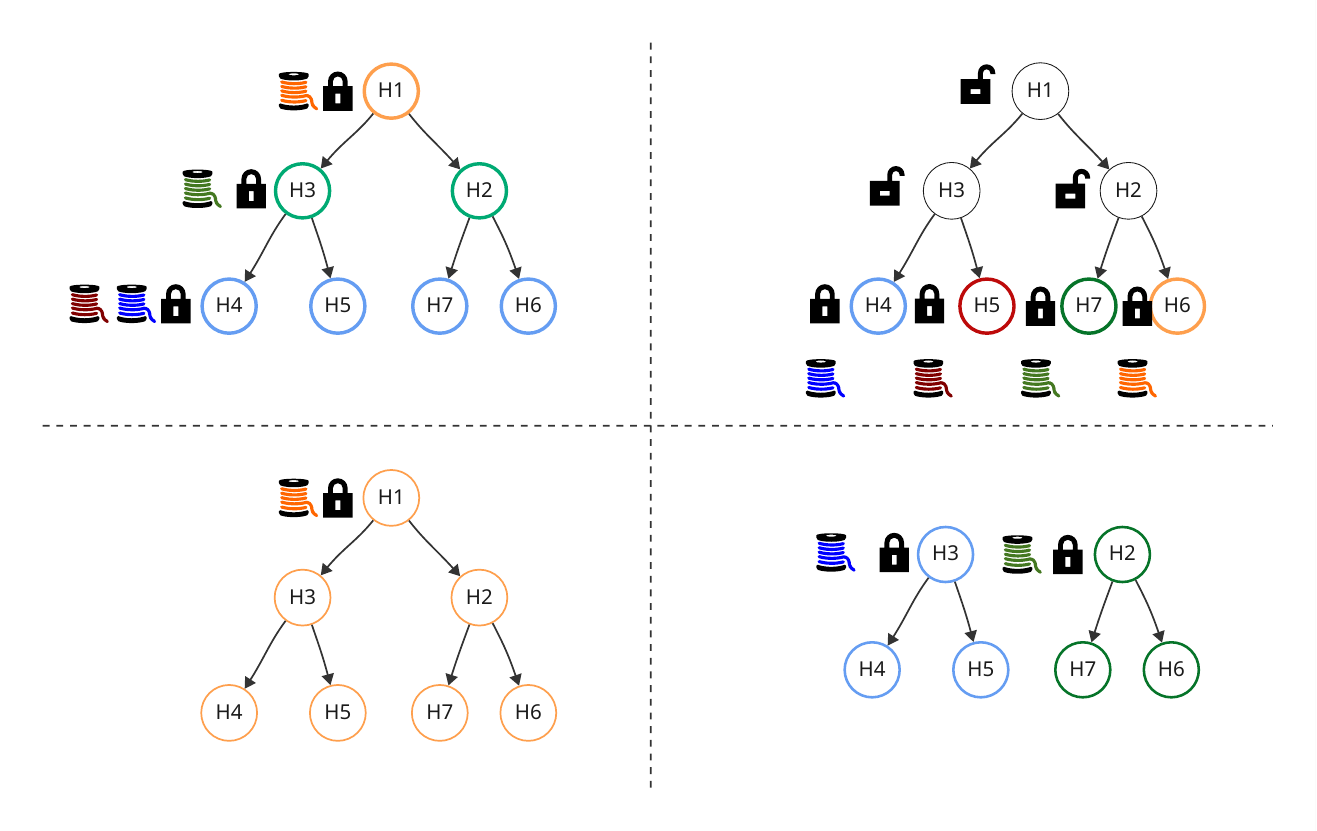}
    \caption{Illustration of the locking and thread allocation. Nodes with an allocated thread are marked in a color associated to the thread.  Lock icons represent exclusive access to a given resource, where open locks represent work that is dynamically allocated.}
  \end{subfigure}

  \caption{A comparison of concurrent Merkle tree implementation families from El-Hindi~\etal~\cite{el-hindiMerkleTreesHighPerformance2023}.  The original figure does not differentiate between compile-time mutual exclusion of thread interference, and run-time mutual exclusion via mutexes.  We have updated the picture to illustrate this point. }%
  \label{fig:el-hindi}
\end{figure}

\point{Characterizing our approach}
El-Hindi~\etal~\cite{el-hindiMerkleTreesHighPerformance2023} elucidate several key challenges commonly encountered when adapting Merkle tree operations for practical use. Firstly, in contrast to traditional data structures, the process of updating information within Merkle Trees necessitates CPU-intensive tasks, such as the computation of cryptographic hash functions or the generation of digital signatures. This results in the performance being fundamentally constrained by the CPU's speed, thereby imposing a limit on the maximum throughput that can be attained. Secondly, attempts to enhance performance through the use of multiple threads that concurrently modify the data structure introduce complexity. Unlike what happens in B-tree variants, this complexity arises because each update requires the root hash of a Merkle Tree to be recalculated, leading to conflicts among threads at the root, which necessitates synchronization to ensure consistency. They focus on an efficient locking scheme for Merkle trees, a variant of a  reverse latch coupling scheme to improve concurrency in Merkle tree batched updates. Moreover, they suggest the concept of splitting to further reduce the contention and the cryptographic overhead at the same time.

They propose a designation scheme according to two properties: amenability to pipelining, and parallelism.  In Figure~\ref{fig:el-hindi}, our approach can be classified as hybrid.  While streaming updates, the computations are done in a lock-free sharded manner.  As soon as the state root is to be computed, our implementation is more similar to the upper-right quadrant.  This allows us to benefit from the greater parallelism during regular operation, while also computing the state root efficiently.

\section{Data Organization}
\label{sec:data-organization}
We begin by describing the data organization in \tool.  Specifically we explain how we arrived at the specific data structures used in both Pleiades and Hyades. 

\subsection{Thread Sharding}
\label{sec:thread-sharding}
For Pleiades, one of the first design goals was to optimize in-memory Merkle tree updates for multi-threaded execution; the reader is referred to other works for more background on this long-standing challenge~\cite{wangExampleParallelMerkle2024,el-hindiMerkleTreesHighPerformance2023,dengAcceleratingMerklePatricia2024}. The outcome we are after is minimizing lock contention. To this end, we note the following:
\begin{observation}
  Because of the inherent integrity-preserving structure, the nodes of the Merkle tree are updated more frequently at higher levels, leading to higher lock contention.
\end{observation}
Given that intra-block observability is optional for most use cases\footnote{Leaf addresses are deterministic, so given $h_k$ we can immediately find the leaf node containing it. }, we point out the following:
\begin{observation}
  The root node only needs to be updated once per block.
\end{observation}

This allows us to partition the key-space into sub-trees with one thread per sub-tree, eliminating the need for mutual exclusion. This also reduces the amount of work that is wasted during repetitious updates. 

We divide the sub-tree space by the number of threads. So each thread knows which subtree it handles.  Specifically, for a~64-core CPU\footnote{SMT provides some speedup, but not one-to-one with a physical core.}, the first six bits of the key determine which thread updates are dispatched to.  The next ten bits determine which sub-tree managed by the thread receives the dispatched state.  This gives us sixteen levels that need neither to be re-balanced, nor updated at every request, but only when the state root needs to be computed.  The precise numbers are determined by the number of cores and can be tuned further manually. The number of levels that do need to be updated is the difference between the bit-depth of the key address space (\textit{e.g.}~24 bits for~16 million keys) and this number~(16) meaning that only \({24 - 16 = 8}\) levels would need to be updated.

Leaf nodes are assigned to a sub-tree statically. This implies that the upper part of the tree has a rigid structure and that the order of computations for generating the state root can be pre-determined.  This relieves us from the need to even instantiate any node structures in memory: computing the hash of the root involves the free-floating hashes of the root sub-tree.  This allows us to further optimize data locality for the final computation and is a small-scale demonstration as to why our data structure does not require re-balancing.  This should be viewed as a continuation of the ideas introduced by El-Hindi~\etal~\cite{el-hindiMerkleTreesHighPerformance2023}.

\begin{figure}[tb]
  \centering
  \includegraphics[width=.75\linewidth]{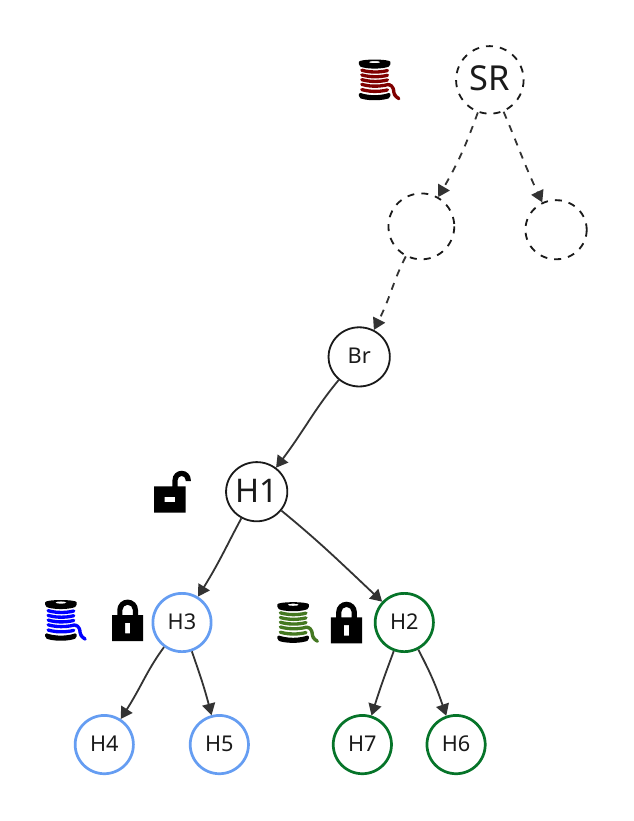}
  \caption{An illustration of the layout of nodes in sparse Merkle tree of \tool. Nodes closer to the state root are drawn to symbolize that they are logical and not instantiated objects.}
  \label{fig:sparse-merkle}
\end{figure}

\subsection{Sparse Merkle Tree}%
\label{sec:sparse-merkle-tree}
The key data structure in the design is the Sparse Binary Merkle Tree.  We chose that authenticated data structure illustrated in \cref{fig:sparse-merkle}, as it is easy to implement, saves memory (as nodes are always fully utilized) and no tree re-balancing is required.  Additionally, this data structure minimizes the size of the state commitment proofs that need to be generated by the historical component.

The Sparse Merkle Tree can be thought of as an ordered collection of \emph{internal} and \emph{leaf} nodes.  The leaf nodes, serving as end-caps of the data structure contain the hashes of keys and values, and represent the account data. Hashing keys generally makes the tree more balanced. Hashing \emph{values} allows reducing memory usage, because the value can then also be stored elsewhere. However, the keys need to be kept:
\begin{observation}
  The keys cannot be omitted, because a \emph{Sparse} Merkle Tree may map several keys to the same leaf node.
\end{observation}
Note, that we are not concerned with collisions after hashing, but collisions between input representations.  On 64-bit architectures all nodes are represented as single 64-bit numbers, meaning that pre-image collisions are much more likely than hash collisions.
% TODO: Clarify value collisions. 

\subsection{Versioning}%
In \tool, version numbers are utilized in both leaf and internal nodes.  Version numbers are an easy way to keep track of updates.  For example, they allow us to store only the updated nodes in a given time period.  For example, the updated nodes from version $x$ to $y$ are all the nodes that have been updated in version $x$, $x+1$, $x+2$, $\ldots$, $y$.  Each version can be stored in its own file, and tracked separately.  By using versions, we simplify the implementation of queries such as ``what was the state diff between version $x$ and version $y$''.

The versioning used here is most similar to versioned data structures in BlockSTM~\cite{blockstm,gelashviliBlockSTMScalingBlockchain2022}.  The main difference is that we version the internal nodes and leaves only.  Since the data is stored append-only in a journal, they are versioned by implication.  Similarly, the version numbers are used as \emph{salt} when computing the hashes of nodes (see Section~\ref{sec:choice-hash-function}).  After a fashion the entire working set can be considered part of a single multi-version data structure, where all the nodes comprising a version represent an \emph{incarnation} borrowing terminology from Gelashvili~\etal~\cite{blockstm}.

% Explaining the content of leaf nodes and branch nodes. 
% TODO: Add figures

This similarity has far-reaching implications.  \tool was designed to work equally well with both pessimistic concurrency control \textit{e.g.} Solana, as well as optimistic concurrency control, \textit{e.g.} Block-STM~\cite{blockstm}.  As of writing, the amount of modifications needed to make \tool{} work with Block-STM is minimal.

A natural consequence of partitioning data by version is that the journal can be split into multiple files by version number.  To control the data sizes of the Merkle tree as well as reduce file contention, we do not store keys and values in-line.  Instead data is written to an append-only journal, and leaf nodes specify the offsets at which the data is stored.

Versioning has one further advantage.  Consider the so-called \(ABA\) update pattern, where a node's value is initially \(A\) then updated to \(B\), and then updated back to \(A\).  With sufficiently fine-grained versioning it is straightforward to preserve the fact that the value has changed. More importantly, state differentials between different versions can be coarse-grained.  One obvious example is coarse-grained for compression for archival purposes.

The final question is when to produce a new version.  While versioning and block production are generally decoupled, there are good reasons to consider tying the two together.  In most blockchains, data between blocks is not observable.  As such, losing information in the \(ABA\) pattern is not a violation of the blockchain logic.  By this logic, in blockchains like Solana the version would be the last \emph{slot} number when the referenced account was changed.  More frequent versions do not change the logic, and are therefore not meaningful.

\section{Optimizations}%
\label{sec:optimizations}

\subsection{Choosing the Hash Function}%
\label{sec:choice-hash-function}

The hash functions are not hard-coded.  One can generically instantiate a Merkle tree supporting any hash function supporting the standard Rust traits.  In the following section we shall explain the main properties that need to be satisfied for an ideal hash function.

To understand why, consider how data is stored.  The ADS is a key-value store, where the key is typically the address of a blockchain entity called an \emph{account}, and is usually, but not always, a cryptographic public key.  One needs to compute the hash of the key \emph{and} the data.  This process is asymptotically linear in the length of the data being hashed. The SHA-family of hash functions need to extend short inputs and perform the same amount of minimum work below a certain length.  Some hash functions can do less work for these short inputs.

\begin{figure}[tb]
  \centering

  \begin{subfigure}{\linewidth}
    \includegraphics[width=\linewidth]{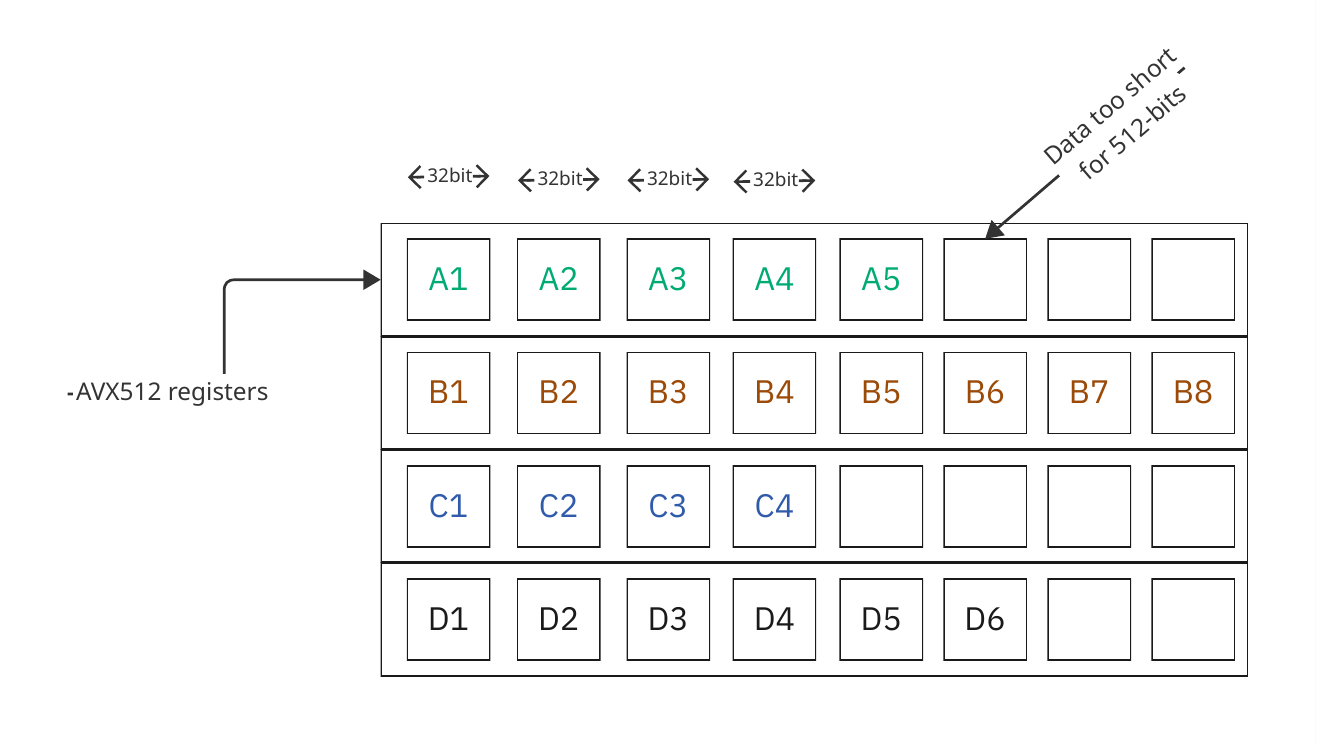}
    \caption{The typical data layout for a latency-optimized hash implementation.}
    \label{fig:simd-layout-straight}
  \end{subfigure}

  \begin{subfigure}{\linewidth}
    \includegraphics[width=\linewidth]{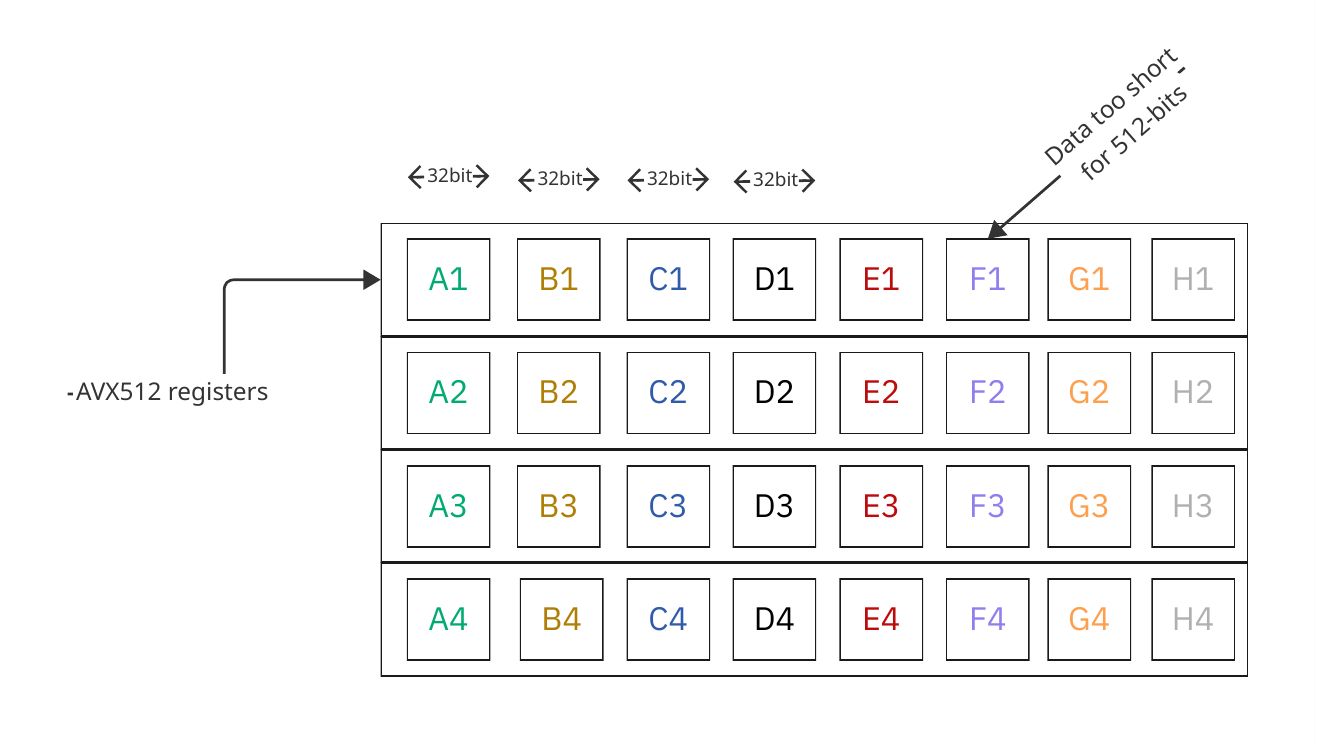}
    \caption{The transposed data layout used for the throughput-optimized hash implementation in \tool. }%
    \label{fig:simd-lanes-transposed}
  \end{subfigure}

  \caption{SIMD layout illustrated.  The inputs are split into chunks (numbered 1 to 8), and designated with Latin letters.  The computations are parallel horizontally, and sequential vertically.}%
  \label{fig:simd-lanes}
\end{figure}

Specifically, if the data is shorter than the SIMD register, by \textit{e.g.}~25\% on average, then~25\% of the work can be avoided with a \texttt{blake2} family of hash function.  Furthermore, if the data is being processed as shown in Figure~\ref{fig:simd-lanes-transposed}, one can intersperse data so that the first chunk of one input can be hashed at the same time as another chunk of another input, minimizing the gaps and increasing the overall degree of parallelism to practically perfect scaling with the width of the SIMD registers.  For a hash function that does not allow the hashing to terminate early for shorter inputs, the final rounds cannot be interspersed, reducing the degree of parallelism, and therefore performance.  It is therefore  a desirable to use a hash function that short-circuits the hash computation for short inputs.

Furthermore, the version number must be encoded in the hash.  While it could be included as data, it is often much more efficient to ``salt'' the hash with the version number.  Allowing salting of hashes is, thus, another desirable property.
Thus, the ideal hash function has the following properties
\begin{enumerate}
\item Pre-image attack resistance, \label{prop:preimage}
\item Reduced work for short inputs, \label{prop:short-circuit}
\item ``Salt'' friendliness, \label{prop:salt}
\item Computational efficiency \label{prop:fast}
\item Optimal hash output width is 256 bits \label{prop:optimal-width}
\end{enumerate}
Thus, SHA-512 and SHA-256 can be considered acceptable by properties~\ref{prop:fast} and \ref{prop:preimage}.  SHA-512 is too long (property~\ref{prop:optimal-width}), in their current implementation the extra bits are truncated, and therefore do not contribute to higher security, explaining our preference for SHA-256.  The Blake2b and Blake2s~\cite{blake2} also offer \ref{prop:salt} and \ref{prop:short-circuit}, while also arguably offering better computational efficiency for similar pre-image attack resistance.  As of writing \texttt{blake2s} is the optimal hash function.

When choosing the hash function, the following are some of the security considerations that we identified during development.
\begin{description}
    \item[Truncation attacks] --- presenting a truncated path terminating at an internal node with the same hash as the expected leaf node. \textbf{Mitigation} --- leaf nodes all inhabit the layer \texttt{0xfff}, unreachable for internal nodes.

    \item[Duplication attacks] --- to foil an exclusion proof for key \(k\) to value \(v\), it is sufficient to include the value \(v\) in the account corresponding to the leaf node's sibling with key \(\kappa\).  \textbf{Mitigation} --- the key is included in the hash of the leaf node.

    \item[Permutation attacks] --- it is possible to produce an invalid inclusion proof by rearranging the siblings: \textit{i.e.} representing a right sibling as a left and \textit{vice versa}. \textbf{Mitigation} --- the branching direction is included in the hash.
\end{description}

%These are by no means an exclusive list of observations; a security audit for any specific hash function should be conducted.

\subsection{Allocation and Prefetching}%
\label{sec:alloc-pref}

The CPU-based implementation of Pleiades faced numerous challenges imposed by the cache hierarchy.  Poor data locality results in suboptimal CPU utilization.  On other hardware platforms, such as GPGPU, these problems are largely dealt with by hiding the latencies,  in the CPU context, the memory layout of the entire database must be tightly controlled to achieve the best performance.

Most programs use the standard, operating-system-provided global allocator.  These often introduce large unpredictable latencies, particularly for allocating large chunks of memory at once, especially given memory fragmentation.  Allocating small objects suffers even greater penalties due to poor data locality.  Briefly, this is because a contiguous memory region can reside completely in cache and often also pre-fetched, while a linked structure can point to arbitrary locations, meaning that every link involves an expensive memory load.

To avoid this, one typically avoids interactions with the global allocator, instead employing what is known as a \emph{slab allocator}.  Specifically, when the program starts a single large contiguous memory region is allocated using the global allocator.  This memory region is then sub-allocated inside the program.  This allows the programmer to optimize the data layout based on the memory access pattern.   For example if one often traverses a tree in depth-first order, one can allocate the left nodes of the binary tree contiguously.

This has further advantages.  In some cases, when the allocation/de-allocation order is known, one can sometimes greatly simplify these operations.  For example, a \emph{bump allocator} can allocate/de-allocate a new object by incrementing/decrementing a single number, which greatly improves performance, at the cost of only de-allocating objects in the reverse order of allocation.  This all results in reduced memory fragmentation data accessed together is allocated together and freed together.

Each Pleiades instance employs two slab allocators: one for nodes and another for leaves, reducing global allocator contention.  The allocators themselves retain a doubly-linked list (to allow allocating dedicated nodes) and embeds metadata into the target type in order to save memory.  We use a doubly linked list to allow allocations to be inserted between two other allocations without traversing the entire list.

Pleiades must support huge trees with billions of entries, which can occupy hundreds of gigabytes of RAM.  This data will live a short amount of time in the CPU cache, which implies that prefetching the data can potentially improve performance significantly. Unfortunately, the (deterministic) access pattern is much too complex for the built-in hardware prefetching to be effective.  Instead, to prefetch a path, we iterate over the bits of the key that are part of the tree, calculating the deterministic address of each node and issuing prefetch instructions for the resulting memory region to get them into the caches.  In our measurements in Figure~\ref{fig:prefetching}, this leads to a performance gains of approximately~60\%.

\subsection{SIMD}%
Single-instruction multiple data extensions are ubiquitous in modern CPUs.  As we have discussed previously, the threads operate on entire sub-trees with no interference, lending to good thread-based scaling.  However, single-instruction multiple data processing is also something that is available to us by virtue of controlling the data layout.

Specifically we found that our implementation with the \texttt{blake2s} hash function, running on an AVX-512 enabled CPU can effectively hash sixteen different branches simultaneously.  This leads to a near-perfect linear speedup of approximately~$16\times$ for a representatively populated tree.
This optimization also makes use of the throughput-optimized hash functions that were designed specifically for \tool.  We call these Throughput Optimized SIMD-friendly Hash functions or TOSH-functions.
\begin{observation}
  Off-the-shelf SIMD-enabled hash functions result in poor utilization of SIMD registers, as they are optimized for the latency of a single hash computation.
\end{observation}
The typical data layout for such computations is illustrated in Figure~\ref{fig:simd-layout-straight}.  Our hash functions, by contrast, are optimized for throughput, by transposing the data as seen in Figure~\ref{fig:simd-lanes-transposed}.  The main advantage of the transposed layout is that data can be mixed filling the registers completely and saturating the SIMD banks.

\subsection{Historical Data via Snapshots}%
\label{sec:snapshots}

Hyades has a different set of performance challenges pertaining to handling historical data.
\begin{observation}
  In almost all high-performance ADS with historical proof support, the main performance bottleneck is disk I/O.
\end{observation}

Because of this, \tool was designed specifically both to work with the historical component disabled altogether, as well as with it enabled and taking periodic snapshots.  This presents a different set of challenges compared to making the historical component being merely faster, because its speed is so significantly determined by the speed of persistent storage and its hardware.  Instead of merely accelerating the historical component, our goal was to minimize the impact of having it enabled.

Figure~\ref{fig:inclusion-proof-physical} illustrates the locations of different elements for an example proof.  The black squares represent the hash and internal entries (not to be confused with internal nodes), the red represents the leaf nodes, which contain their data payloads in the journal, and blue nodes represent external entries for different versions (not used in the inclusion proof).
\begin{observation}
  The version space of~52 bits is enough to support a new version every microsecond for the next~100 years.
\end{observation}
At this point data is written to persistent storage.  It can be unpacked for easier proof generation, or it can be archived to reduce storage footprint.  Within these constraints, 52-bits provides sufficient headroom.
\begin{observation}
  52-bit numbers can address files up to~4 petabytes in size.
\end{observation}

\point{Snapshot frequency} The communication between the in-memory and historical components happens through snapshots (see Figure~\ref{fig:arch}).  As such, the snapshot creation frequency is an important tunable configuration parameter that should be selected based on the underlying hardware: \textit{e.g.} how many CPU cores, how many disks, what are the speeds of these components, software, latency budgets imposed by the protocol: \textit{e.g.} Solana blocks are produced every~400ms; as well as other factors.

We believe that the number of factors is so large that it is preferable to set these parameters empirically: measuring the overall system throughput under synthetic load and adjusting the snapshot frequency to increase the throughput.

% TODO: Compare to QMDB proof system, show why our proof generation is simpler

\section{Proof System}%
\label{sec:proofs}

The primary goal of Hyades is to generate inclusion and exclusion proofs from snapshot files.  Algorithm~\ref{alg:gen_algo}, allows the information to be encoded much more efficiently than the naive linked structure, at the cost of some complexity in Hyades.

In summary, generating a proof involves identifying which sub-tree a key would belong to.  This is done straightforwardly as the sub-tree is determined from the hash of the key.  Next, the sub-tree is traversed in a specific order which is again determined by the hash of the key.  At this point one of two things may happen.  Either the node exists and the DB can serve an inclusion proof, \textit{i.e.} the path to the corresponding leaf and the content of the leaf node, or the node does not exist and the DB serves an exclusion proof, \textit{i.e.} the path that was traversed to determine the absence of the leaf. In both cases, the requester can then re-hash the path from bottom to top to prove the inclusion/exclusion of the leaf.

As such, the data structure described in Section~\ref{sec:sparse-merkle-tree} requires the following form of state commitment proof.  To provide an inclusion proof one must provide the location of the leaf node where a given key is set to a given value, but also prove that this information is consistent with the rest of the state of the Merkle tree.

\subsubsection{Example: Inclusion Proof}%
\begin{figure}[tb]
    \centering
    \includegraphics[width=\linewidth]{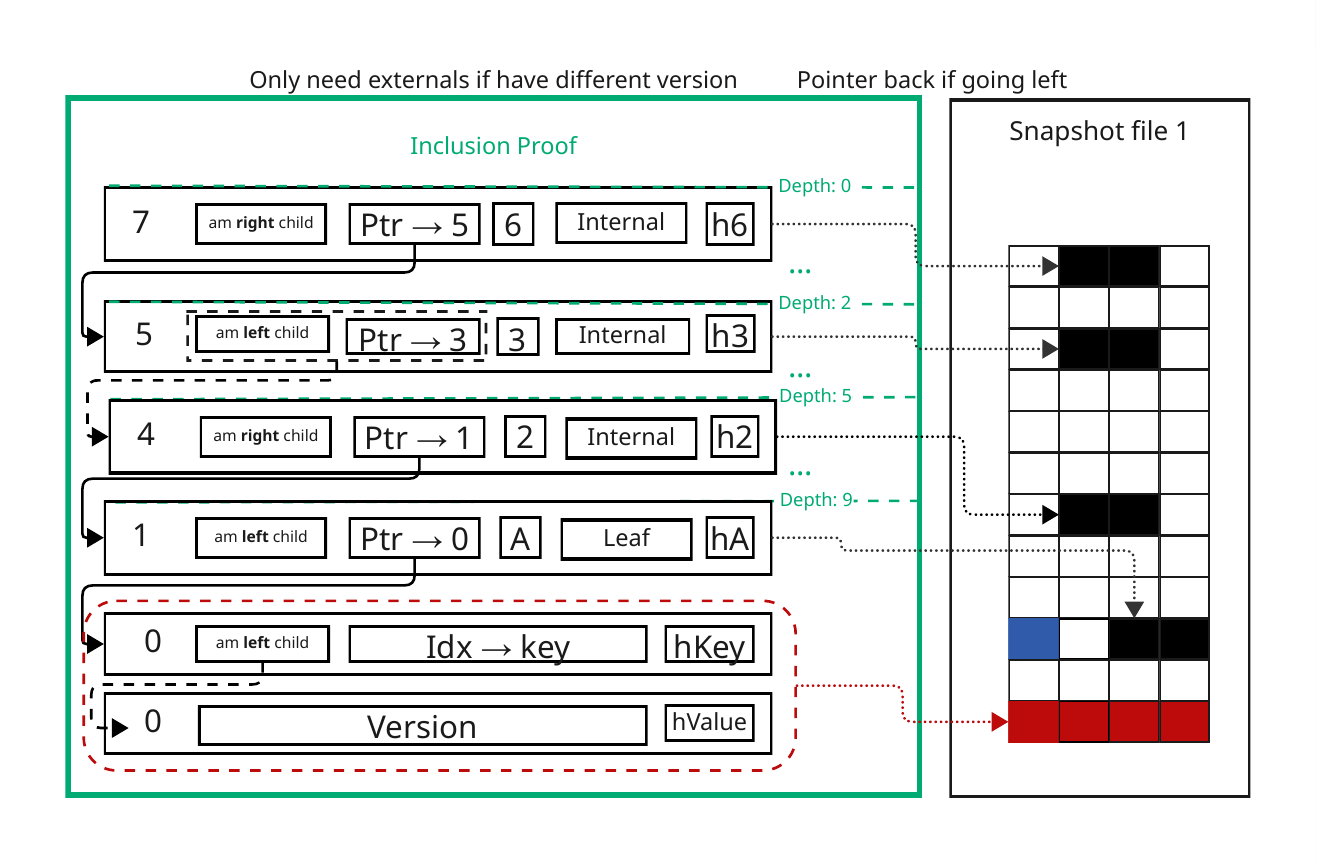}
    \caption{Data layout of the inclusion proof. Boxes indicate entries with their labels. The dotted arrows represent the file locations in the snapshot file. The dashed lines represent logical connection, while solid lines physical presence.}%
    \label{fig:inclusion-proof-physical}
\end{figure}

\begin{figure}[tb]
  \centering
  \includegraphics[width=.5\linewidth]{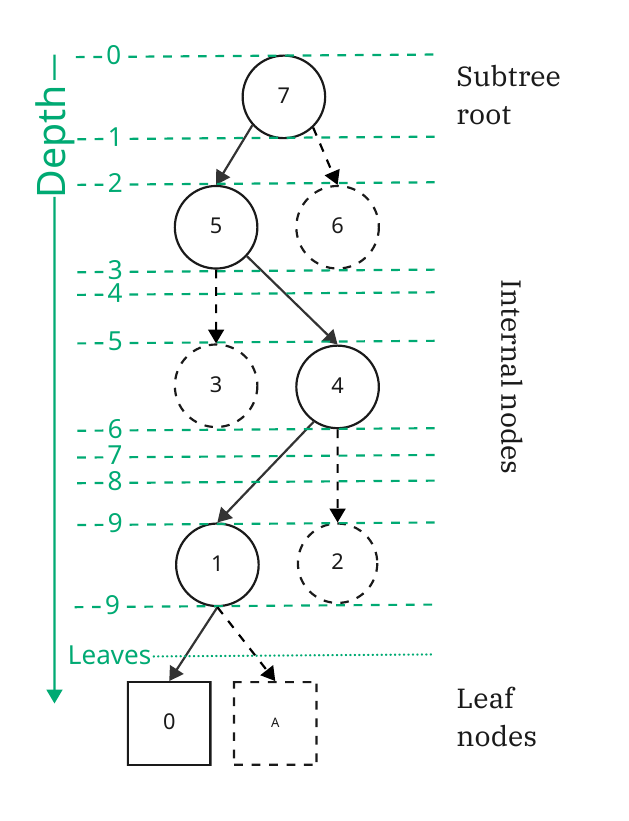}
  \caption{A logical representation of the Merkle proof as a linked structure. The solid lines represent physically instantiated objects, such as node references, while dashed lines represent information that must be inferred.}
  \label{fig:inclusion-proof-logical}
\end{figure}

The logical process is best seen with an example, presented in Figures~\ref{fig:inclusion-proof-logical} and~\ref{fig:inclusion-proof-physical}.  The proof shown here terminates at the sub-tree root, which for our purposes is the same as terminating at the overall state root.  The depth of the nodes is to be encoded, because we are operating on a \emph{Sparse} Merkle tree.  The data to be verified is the key \texttt{key} corresponding to value \texttt{value}, stored in the leaf node~\Circled{0}.

First we traverse from node~\Circled{7} to node~\Circled{5}, to node~\Circled{4} to node~\Circled{1} and finally to node~\Circled{0}.  At each level we check if the depth agrees with what we anticipate.  Do we have an internal node where we expect a leaf?  Do we get a right child instead of a left child.   Because the hashes of the nodes on the main path: \({\Circled{7} \rightarrow \Circled{5}}\), \({\Circled{5} \rightarrow \Circled{4}}\), \({\Circled{4} \rightarrow \Circled{1}}\), \({\Circled{1} \rightarrow \Circled{0}}\) are to be verified, there is no point in providing them.  Instead, once we reach the leaf node~\Circled{0}, we start evaluating the expected hashes of the nodes in reverse order.  Specifically, we evaluate
\begin{align}
  h_{0} = &H(h_{\text{Key}} | h_{\text{Value}}) \\
  h_{1} = &H(h_{0}|h_{A}) \\
  h_{4} = &H(h_{1} | h_{2}) \\
  h_{5} = &H(h_{4} | h_{3}) \\
  h_{7} = &H(h_{5} | h_{6}) \label{eq:h7}
\end{align}
where \(H\) is the hash function,  \({h_{0} \ldots h_{7}}\) are the hash values of the nodes.  \(h_{\text{Value}}\) and \(h_{\text{Key}}\) as well as \(h_7, h_6, h_3, h_2, h_A\) are provided in the inclusion proof and we must verify that the value obtained via \ref{eq:h7} matches the one that we were provided in the inclusion proof.

The schematic representation of the physical data layout which corresponds to the example is shown in Figure~\ref{fig:inclusion-proof-physical}.  The snapshot typically contains multiple inclusions proofs laid out to minimize data storage, as such, the precise locations of all objects are indicated with an offset inside the journal file.
The main unit is an entry.
\begin{enumerate}
\item \textbf{Internal pointers:}
The internal pointers are directions to a node that is stored in the current file.  They comprise a \emph{tag} (which is an offset inside the file), two bits indicating whether this is right child and if this is a leaf node, and a byte representing the depth in the tree at which the branching is made.

\item \textbf{External references:}
The external references represent nodes that are in a different file.  They contain a \emph{tag} representing a version, the same bit flags, and an indication of whether or not this specific field is a right child.

\item \textbf{Key:}
Which is an offset inside of the file to the location where the key is placed.

\item \textbf{Leaf:}
Which is a version for which that particular leaf is defined.
\end{enumerate}
Each entry is preceded by a hash~\ref{fig:inclusion-proof-physical}, which usually describes the \emph{following node} not the current one, with the exceptions of key and leaf nodes. So, for example~\Circled{7} contains the hash of node~\Circled{6}.

Entries can be viewed as a structure encoding information about two child nodes and the depth at which this entry's corresponding node resides.  The internal reference entry encodes the location of the next \emph{main} child and whether or not the \emph{current} node is a left or a right child.  For example, the node~\Circled{1} is a left child of node~\Circled{4}, and contains the pointer\footnote{We differentiate pointers, from offsets/indices because they have different semantic meaning and in our current implementation different possible ranges.  In effect they both represent an offset into the snapshot file, but belong to different kind of entries, so pointers only appear in internal reference entries} to  node~\Circled{0}.

Because of the depth-first order of traversal, the pointer doesn't always point to the correct main child, \emph{left} children are handled differently from \emph{right} children, as can be seen with \({\Circled{5} \rightarrow \Circled{4}}\) the pointed-to node is actually~\Circled{3}, the location of node~\Circled{4} we \emph{infer} from the memory layout of the Merkle tree\footnote{Specifically, the Merkle tree is flattened in memory, meaning that siblings' locations can be computed with pointer arithmetic.}.  Internal entries must also indicate the hash of the \emph{sibling} of the main child, and whether or not that node is internal.  This is used for salting the hash to prevent permutation attacks.  The tree is sparse, so we must include the depth of nodes as actual data stored in-line (indicated with green solid lines).  The depth of all leaf nodes is \texttt{0xfff} as explained in Section~\ref{sec:choice-hash-function}.  Internal entries are strongly typed, and as such one must indicate whether the next main node is a leaf.

The ``leaf'' entries and ``key'' entries are separate entities to preserve the uniform sizes of entries.  They always appear together and carry complementary information, one carries the version and the offset to the value, the other carries the hash of the key and the key itself.  As such an inclusion proof will typically comprise several internal references (4 in our example) and two entries representing the leaf node.

The entries are packed and fit into~64-bits on~64-bit architectures\footnote{and~52 bits on~32-bit ones}.  When unpacked, the Entries occupy three times as much space.  This allows for excellent cache-friendliness and essentially make the proof generation\footnote{The same goes for verification, where cache-friendliness allows the entire proof to be verified with minimal memory loads, thus optimizing for latency.} a memory-bound operation.

\subsubsection{Snapshot Anatomy}
The snapshots are the only persistent storage.  The snapshot files contain segregated representations of the in-memory tree, (which we shall use to construct a proof in Section~\ref{sec:proofs}), as well as an append-only journal, representing the data added in the current version\footnote{Data from different versions is encoded as an external reference.}.

To achieve this, each proof is composed of several entries. Each entry encapsulates a compact~40-byte value containing the leaf's value, key, as well as references to existing internal nodes or external versions within the file.  As such, an \emph{entry} can be thought of as a tagged union of the following:

%
%This structure is generated with the following pseudocode:
{
\footnotesize
\begin{algorithm}
  \SetKwData{SubtreeRoots}{subtree roots}
  \SetKwData{Leaves}{leaves}
  \SetKwData{Nodes}{nodes}
  \SetKwData{Version}{version}
  \SetKwData{Buffer}{buf}

  \SetKwFunction{Save}{save}
  \SetKwFunction{Offset}{Offset}
  \SetKwFunction{Internal}{internal}
  \SetKwFunction{External}{external}

  \Leaves: A slab containing leaf nodes

  \Nodes: A slab containing internal nodes

  \Version: Externally set version

  \Buffer: The output buffer to be written to file

  \BlankLine

  \For{$n \gets \SubtreeRoots$}{
    $l, r \gets n.children$\;
    $d \gets n.depth$\;
    \(h_{l}, h_{r} \gets (l, r).hash\)\;
    \(V_{l}, V_{r} \gets (l, r).version\)\;

    \uIf{$V_l = \Version$} {
      $\Save{l,  \Leaves, \Nodes,  \Version, \Buffer}$\;
      \eIf{$V_r = \Version$} {
        $m \gets \Offset{\Buffer.current\_offset}$\;
        $\Internal(0, V_{r}, h_{r}, isRight, d, \Buffer)$\;
        $\Save(r, \Leaves, \Nodes, \Version, \Buffer)$\;
        $\Internal(m, V_l, h_l, isLeft, d, \Buffer)$\;
      } {
        $\External(V_r, h_r, isRight, d, \Buffer)$\;
      }
    }\uElseIf{$V_r = \Version$} {
      $\Save(r, \Leaves, \Nodes, \Version, \Buffer)$\;
      $\External(V_l, h_l, isLeft, d, \Buffer)$\;
    }\Else {
      $\External(V_r, h_r, isRight, d, \Buffer)$\;
      $\External(V_l, h_l, isLeft, d, \Buffer)$\;
    }
  }
  \caption{Generating the snapshot file.} 
  \label{alg:gen_algo}
\end{algorithm}
}

Algorithm~\ref{alg:gen_algo} shows how one would construct the state. Much of it has to do with deciding whether to include an internal or external entry.  The order in which the nodes are traversed, and appended is strict, and determined by locality requirements.

\subsection{Constructing a Proof}
\label{sec:construction}
% Consider moving this
In this section we would like to explain
\begin{enumerate}
\item How to construct an inclusion/exclusion proof for an arbitrary key \texttt{key} and value \texttt{value}, given a snapshot file;
\item How this procedure is different from \textit{e.g.} the corresponding procedure in QMDB~\cite{qmdb}
\end{enumerate}
The construction algorithm was outlined in the beginning of Section~\ref{sec:proofs}: find the sub-tree, traverse it, find a leaf (or not) check the hash and output.  In this section we describe it in much more detail.
{\footnotesize
\begin{algorithm}
  \SetKwData{None}{None}
  \SetKwData{STDIN}{STDIN}
  \SetKw{Type}{type}
  \SetKw{Break}{break}
  \SetKw{Continue}{continue}
  \SetKw{Is}{is}
  \SetKwFunction{Some}{Some}
  \SetKwInOut{Input}{Input}
  \SetKwInOut{Output}{Output}

  \Input{Key \(k\) and \(entries\)}

  \textbf{where}

  \(h_{k} \gets H(k)\): The hash of key\;
  \(entries\): Collection of entries from snapshot file\;

  \BlankLine
  \Output{list of entries \(Path\), \(external\) node, \(k_{f}\) and \(v_{f}\), and position \(P\)}
  \textbf{where}

  \(Path \gets []\): The path of nodes traversed

  \(external \gets \None\): The external entry references

  \(k_{f}, ver_{f} \gets \None, 0\): The \emph{final key} and \emph{final version}

  \(P \gets entries.len - 1\): The final \emph{cursor position} in the file

  \BlankLine
  \While{\(true\)} {
    \(item \gets entries[P]\)\;
    \Switch{\(item.info\)} {
      % TODO: These should look more like \If
      \(\mathbf{Key} \mapsto (k_{f}, P) \gets (H(item), P-1)\)\;

      \(\mathbf{Leaf(v)} \mapsto ver_{f} \gets \mathbf{v}\) \(\Break \)\;

      \(\mathbf{Internal(ptr)} \mapsto \)

      \(d \gets \mathbf{ptr}.depth\)\;
      \eIf{\(\left\{h_{k}[d/8] \& \left[ 1 \ll (d \mod 8) \right] \ne 0\right\}  = \mathbf{ptr}.isRight\)} {
        \(P \gets \mathbf{ptr}.tag\)\;
        \(\Continue\)\;
      } {
        \(Path.push(item.hash, \mathbf{ptr}.isRight, d, 0)\)\;
        \(P \gets P - 1\)\;
      }
      \(\mathbf{External(ptr)} \mapsto\)
      \If{ \(external = \None\) }{
        \(d \gets \mathbf{ptr}.depth\)\;
        \If{\( \left\{ h_{k}[d/8] \& \left[ 1 \ll (d \mod 8) \right] \ne 0 \right\}  = \mathbf{ptr}.isRight\)} {
          \(external = \Some(\mathbf{ptr}.tag)\)\;
        }
      }
      \If{\(Path.last().depth = \Some{d}\)} {
        \Break
      }
      \(Path.push(item.hash, \mathbf{ptr}.isRight, d, \mathbf{ptr}.tag)\)\;
      \(P \gets P - 1\)\;
    }
  }
\caption{The traversal of the proof.}%
\label{alg:traversal}
\end{algorithm}
}
Algorithm~\ref{alg:traversal} indicates the work that one needs to do to construct the path.  Let's start by noting that the key only matters in so far as its hash can be computed \(h_{k}\).  During the loop iterations, we update the \emph{final key} \(k_{f}\) and the \emph{final version} \(ver_{f}\) based on whether we identify a \(\mathbf{Key}\) node or a \(\mathbf{Leaf}\) node carrying the version \(\mathbf{v}\).  These can be regarded as the base case, since finding the \(\mathbf{Key}\) entry guarantees finding the \(\mathbf{Leaf}\) entry next, which terminates the loop.

The two other cases are more interesting.  An \(\mathbf{Internal}\) entry carrying an \emph{internal pointer} \(\mathbf{ptr}\) encodes a considerable amount of information.  First, there's the \(\mathbf{ptr}.depth\) which indicates the level of the sparse Merkle tree that we are currently in.  Based on the depth \(d\) the hash of the key \(h_{k}\), as well as whether or not the current node is a ``right'' child, the path is chosen.  The path can either be following the pointer's tag: \({P \leftarrow \mathbf{ptr}.tag}\) but it can also be traversing to the adjacent entry: \({P\leftarrow P-1}\).  External references follow the same logic with regard to traversing the path, but also offer another means of terminating the loop, specifically a situation in which depth of the last path entry, as well as the current depth match.

This is the algorithm by which one would construct an inclusion or exclusion proof path, such as the example in Figure~\ref{fig:inclusion-proof-logical} given a snapshot file provided as a series of \(entries\).  However to verify this proof (and to determine whether it's a proof of inclusion or exclusion) one has a small amount of outstanding work that needs to be performed.

\subsection{Verifying a Proof}%
\label{sec:verifying-proof}
% This information MUST BE PRESERVED; DO NOT REMOVE,
% WRAP in \begin{comment}

In a Merkle tree a proof (of inclusion or exclusion) is a string of nodes terminated by the state root on the one end and the leaf node with the key and value on the other.  To verify an inclusion proof is to ascertain that
\begin{enumerate}
\item Every node included in the proof is also in the Merkle tree provided to the prover;
\item Every node has exactly the same hash as in the Merkle tree provided by the prover;
\item (For inclusion) the key and the value hashes reproduce the state root;
\item (For exclusion) the key and the value hashes lead to a contradiction (\textit{i.e.} the chain of nodes where the data would have been is not there)
\end{enumerate}
To verify the proof one needs to traverse back along the \(Path\) in first-in/first-out traversal in Algorithm~\ref{alg:verification}.  Along that path we compute what will be the root hash \(h_{\text{root}}\) for the overall tree by using the property of the Merkle tree and combining the peer hash \(h_{\text{peer}}\) with the root hash.  The attentive reader will have noticed that we need to take into account whether a node is a right or left sibling, to prevent permutation attacks.  Furthermore, we are salting the hash with a value derived from the version \(ver_{f}\) and the depth \(depth\), where leaf nodes have an implicit depth of \texttt{0xfff}.  If at the end of the process, we obtain the hash of the final key agreeing with our input \(h_{k} = H(k)\), then the overall proof is a proof of inclusion.  If we do not, it is a proof of exclusion.

{\footnotesize
  \begin{algorithm}
    \SetKwInOut{Input}{Input}
    \SetKwInOut{Output}{Output}
    \SetKwFunction{Salt}{Salt}
    \SetKwFunction{Some}{Some}

    \Input{Version \(ver_{f}\), entries \(entries\), cursor position \(P\), path \(Path\) and final key \(k_{f}\), \(external\)}
    \BlankLine
    \(h_{\text{root}} \gets H(\Salt(ver_{f}, 0xfff), [k_{f}, entries[P].hash])\)\;
    \While {\(\Some(h_{\text{peer}}, isRight, depth, ver_{c}) \gets Path.pop()\)} {
      \(ver_{f} \gets \max [ver_{f}, ver_{c}]\)\;
      \(hash\_inputs = \) \eIf {\(isRight\)} {[\(h_{\text{root}}, h_{\text{peer}}\)]} {[\(h_{\text{peer}}, h_{\text{root}}\)]}

      \(h_{\text{root}} \gets H(\Salt(ver_{f}, depth), hash\_inputs\)\;
    }
    \uIf{\(\Some(ver) = external\)} {
      proof version \(ver\) for \(h_{k}\)\;
    } \uElseIf{\(h_{k} = H(k_{f})\)} {
      proof inclusion for \(h_{k}\)\;
    } \Else {
      proof exclusion for \(h_{k}\)\;
    }
    \caption{Verifying the proof.}%
    \label{alg:verification}
  \end{algorithm}
}

% TODO: Expand arrows in fig:arch

% Local Variables:
% jinx-languages: "en_US"
% TeX-master: "main"
% End:

\section{Experimental Results}%
\label{sec:eval}

\subsection{Scaling Pleiades}
All results below were produced with the Pleiades-bench harness compiled under Rust \texttt{nightly-2025-06-20}.  The workload comprises 256-bit uniformly random keys (no cache-friendly prefixes), a mix of updates~(90\%), inserts~(5\%), and deletes~(5\%).  The number of leaves was at least~128M to keep the working set far outside the L3 cache of this specific CPU.

For our measurements, we rented an AWS i7ie-metal-48xl server machine.  It comes with two Xeon~8559C CPUs with overall~96 cores and~192 threads.  It is equipped with~1,536~GB RAM as \(16\times\) 96GB~DDR5--5600 and has \(16\times\) AWS Nitro v3 NVMe SSDs running in Raid~0 and formatted with the \texttt{xfs} file system.

\subsubsection{Line-Speed Processing}
Given the asynchronous nature of \tool's approach, we primarily concern ourselves with the throughput the system delivers, as opposed to latency.

Solana has roughly one billion accounts active and otherwise.  Furthermore, many of the results reported by authenticated database benchmarks investigate the one billion to eight billion key range.  Thus we have chosen to provide benchmark numbers at~227 (128  million),~230 (one billion) and 233 (eight billion) accounts.

Under the~1B key setting, the~96-core AWS machine sustained over~48M state updates  per second without Hyades, and even peaked at~60M updates per sec, a near-perfect scaling result, with each CPU core contributing approximately~0.5M state updates per second.  Single-core runs peak at~0.64 million updates per second, so end-to-end efficiency sits at roughly~78\% of the isolated core rate.

\subsubsection{NUMA and Memory Bandwidth Constraints}
On a dual-socket system with a eight-channel DDR5 bus, cross-socket latency and finite memory bandwidth prevent each core from reaching its solo peak. These NUMA effects and bus saturation cap per-core contributions at~78\% in full-system runs.

Enabling SMT additionally confirms that the memory-latency is the bottleneck here: hyper-threads improve throughput by~40\%.  As one thread is waiting on memory, the other can already do useful work.  We  validated  this thesis by using the platform-provided hardware monitoring tools and identifying the longest CPU pipeline stalls as caused by DRAM access.

Enabling historization via Hyades cuts the performance in half to~24M~ups.  This is the number that would be fair to use when comparing with other projects, such as QMDB, as historization is an integral part of their implementation.  We report both numbers, as our architecture allows us to have history on a separate machine.

Each update serializes~400 bytes of snapshot data (leaf segment plus inner-node hops).  If we dedicate~64 cores and 128-threads  of the CPU of the host server to the the snapshot stream, this corresponds to a total of~50~Gbps, roughly~50~MB/s per thread, well within the burst capacity of a two PCIe $4\times 4$ NVMe drives, sustainably handled by four commodity SSDs, and easily accommodated by a~50~GbE link.  We remind the reader that our\footnote{\url{https://www.eclipselabs.io/blogs/breaking-10-million-tps}} ed25519 signature verification software can (on a 64-core server) similarly saturate that specific line rate.

\begin{figure}[tb]
    \centering
    \includegraphics[width=\linewidth]{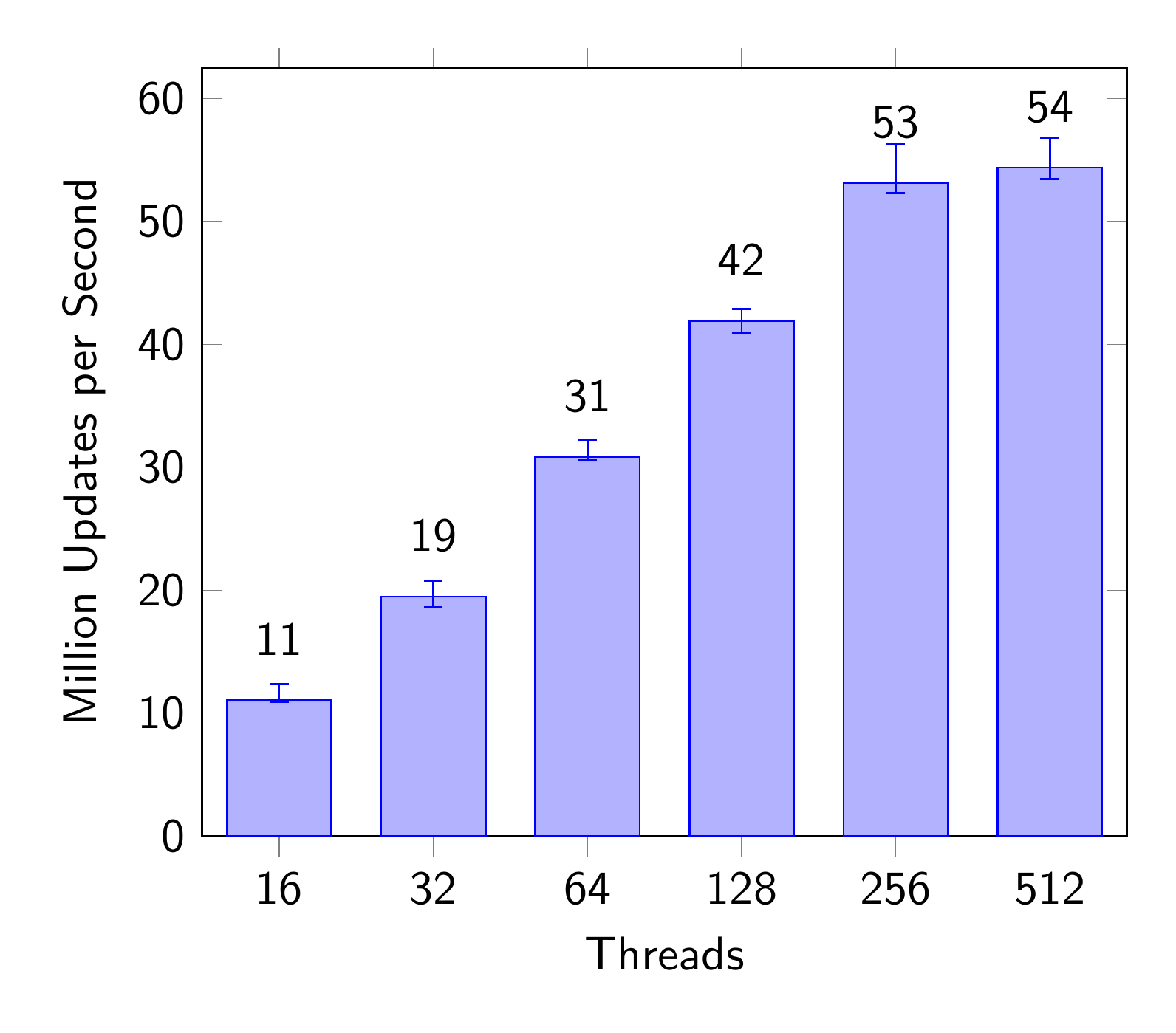}
    \caption{Scaling with number of worker threads.}%
    \label{fig:worker-threads}
\end{figure}

\begin{figure}[tb]
    \centering\footnotesize
    \setlength{\tabcolsep}{2pt}
    \begin{tabular}{p{1.2cm}p{2cm}p{2cm}p{2cm}}
        \toprule
        Commit period (ms)	& \tool{} throughput at $2^{27}$ accounts &	Speedup compared to QMDB v0.2.0	& Speedup compared to FAFO/QMDB\\
        \midrule
        500	& 24,100,000	& $19\times$	& $30\times$\\
        100  & 14,800,000	& $11\times$	& $18\times$\\
        \bottomrule
    \end{tabular}
    \caption{Comparing \tool{} to QMDB and FAFO.}%
    \label{fig:comparison-to-qmdb}
\end{figure}

With a fixed 128M key space, we varied the number of worker threads handling the updates from 16 up to 512.  Throughput rises sub-linearly until the dual-socket memory subsystem saturates.  This confirms that Pleiades is memory-bound.

We observe that root hash generation, historically the slowest stage of block production, now occupies a single-digit share of CPU time.  For Eclipse, this effectively eliminates state commitment computation as a throughput bottleneck, being able to operate at line speed on fast networks.  Furthermore, our implementation of Pleiades does not require a special kernel or accelerator card.

\subsubsection{Throughput vs.  Key‐Space Size}
Our design aims to provide eventual user address space growth.  We measured sustained state updates per second over a~10-minute run with three key-space sizes:~128M,~1B, and~8B accounts; the latter could be considered to be overkill.  Throughput stabilizes as the store is filled and remains stable thereafter. At~1 billion accounts, we get more than~45M state updates per second, which would correspond to over~15M TPS.\@ Even at~233 accounts (over~8B), well beyond typical chain sizes, the engine sustains nearly~40M state updates per second.

\begin{figure}[tb]
    \centering
    \includegraphics[width=\linewidth]{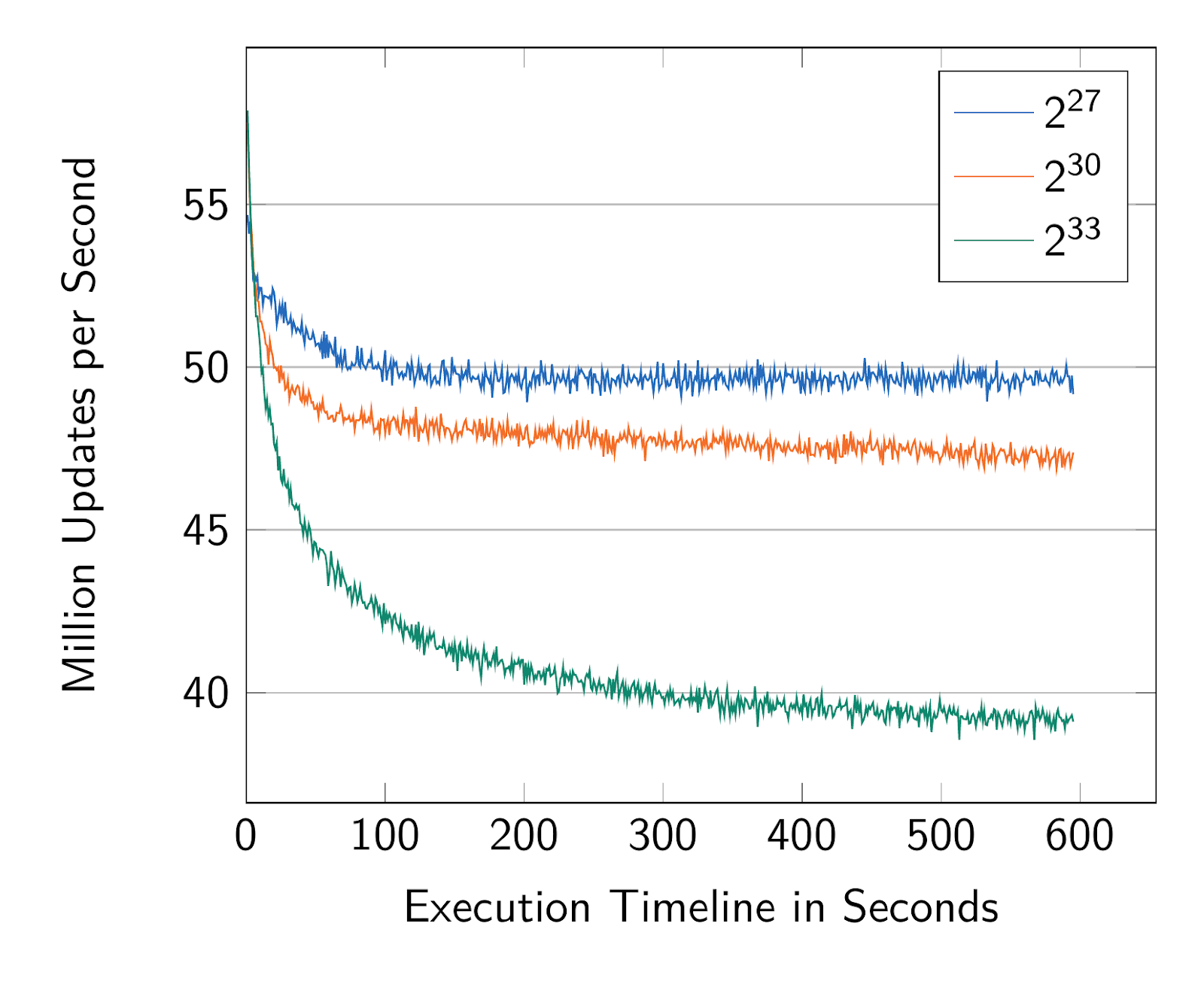}
    \caption{Scaling with Key Sizes}%
    \label{fig:key-size}
\end{figure}

\begin{figure*}[tb]
    \begin{subfigure}{0.45\linewidth}
    \includegraphics[width=\textwidth]{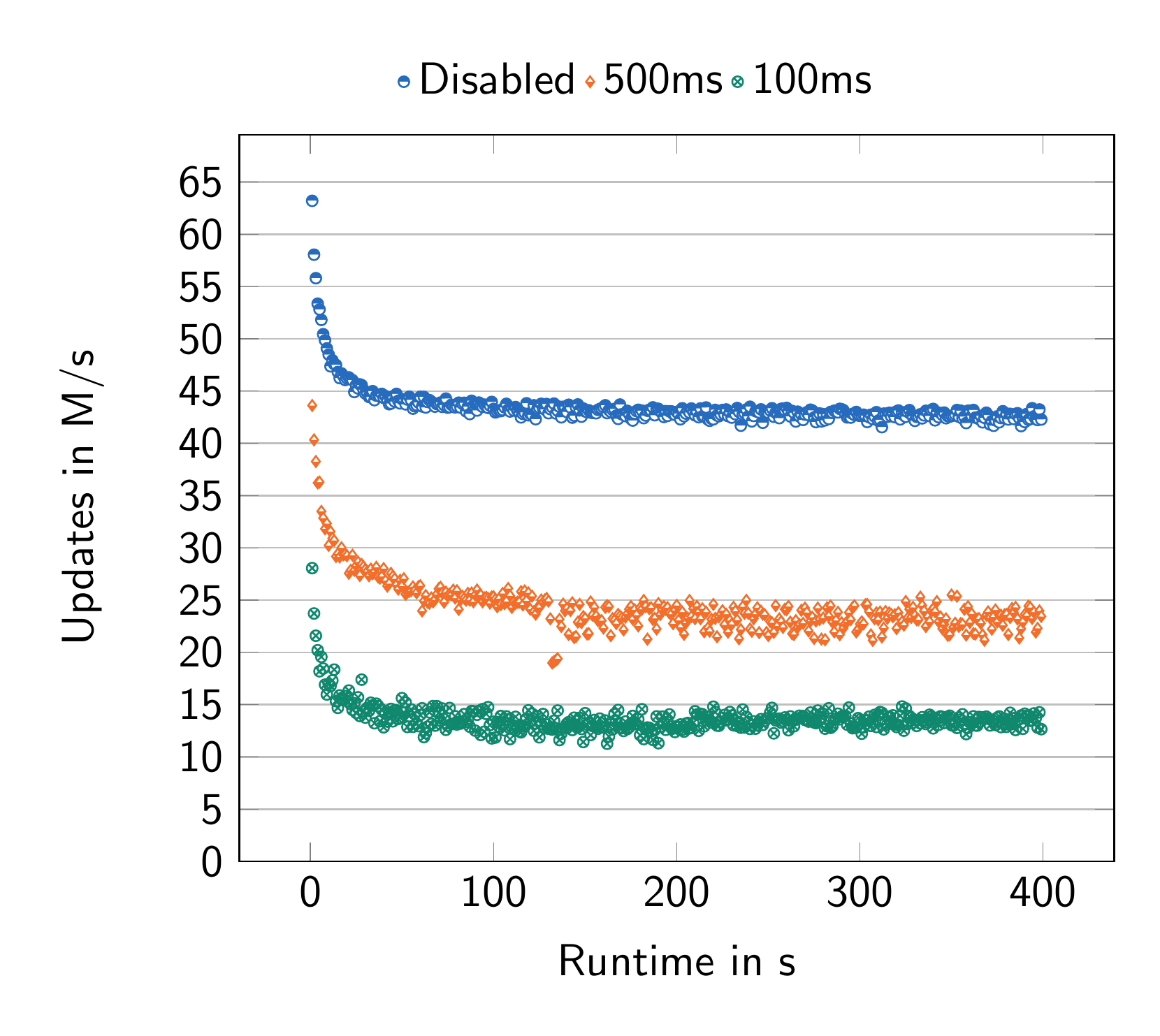}
    \caption{Measured on an AWS Intel Sapphire Rapids system.}%
    \label{fig:snapshot-intel}
    \end{subfigure}\hfill
    \begin{subfigure}{0.45\linewidth}
    \includegraphics[width=\textwidth]{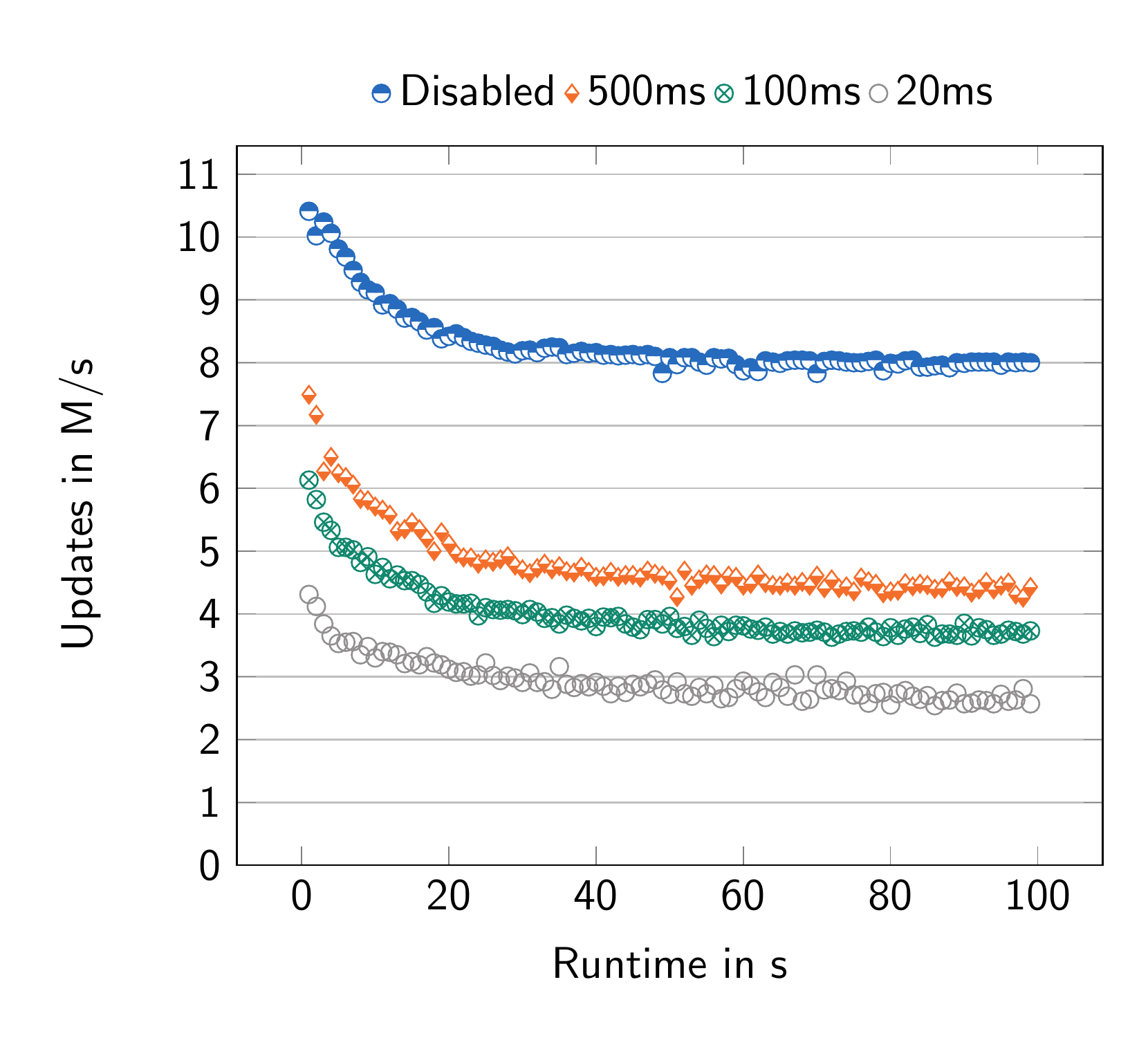}
    \caption{Measured on an Apple M4 Macbook Pro.}%
    \label{fig:snapshot-macbook}
    \end{subfigure}
    \caption{The effects of snapshot period on runtime performance.}%
    \label{fig:snapshot-period}
\end{figure*}

\subsection{Hyades: Snapshots and History}
\subsubsection{Effects of Snapshot Frequency}
When first designing \tool, it was apparent that the historical state shall be the major bottleneck.  It was also evident that only a subset of the operations required the historical data.  This motivated our separation of the historical data store~---~Hyades,  and the in-memory data store~---~Pleiades.  This would allow workloads that need an as-fast-as-possible state commitment (re)-computation to operate without the overhead of interacting with (typically slower) persistent storage.

Figure~\ref{fig:snapshot-period} demonstrates the direct impact of snapshot period on overall system throughput.  For ease of reproduction, measurements were additionally conducted on an Apple M4 Pro machine (Figure~\ref{fig:snapshot-macbook}), obviating the need for server deployment on AWS.\@ Throughput reaches its maximum at~8M updates per second with snapshots disabled.  A~500ms snapshot period reduces throughput by a factor of two, consistent with the observed decrease from 48 million to~24M updates per second on the~96-core AWS machine (Figure~\ref{fig:snapshot-intel}).

This design allows the protocol to retain flexibility.  For example, machines that cannot keep up with the chain at the~400ms cadence, could keep up with the chain at the~800ms cadence, or at another integer multiple of~400ms.  These nodes can still produce useful work, the snapshots produced at the~800ms mark should be identical across the slowest nodes.

\begin{figure}[tb]
    \centering
    \includegraphics[width=\linewidth]{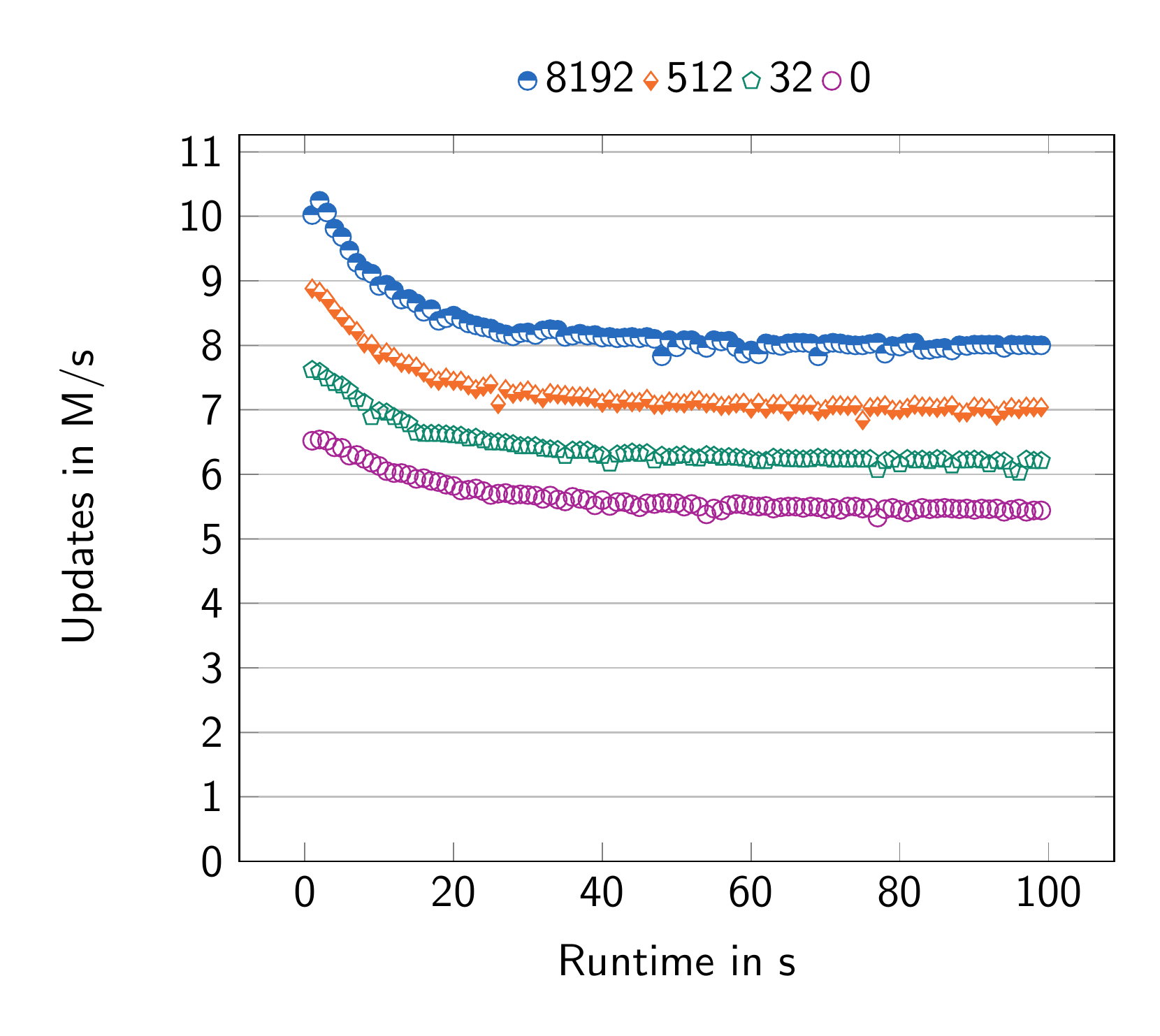}
    \caption{The effects of varying the number of subtree roots on the Pleiades (in-memory data storage) throughput.  Measured on the Apple M4 Macbook Pro}%
    \label{fig:varying-subtrees}
\end{figure}

\begin{figure*}[tb]
  \begin{subfigure}{0.3\linewidth}
    \includegraphics[width=\textwidth]{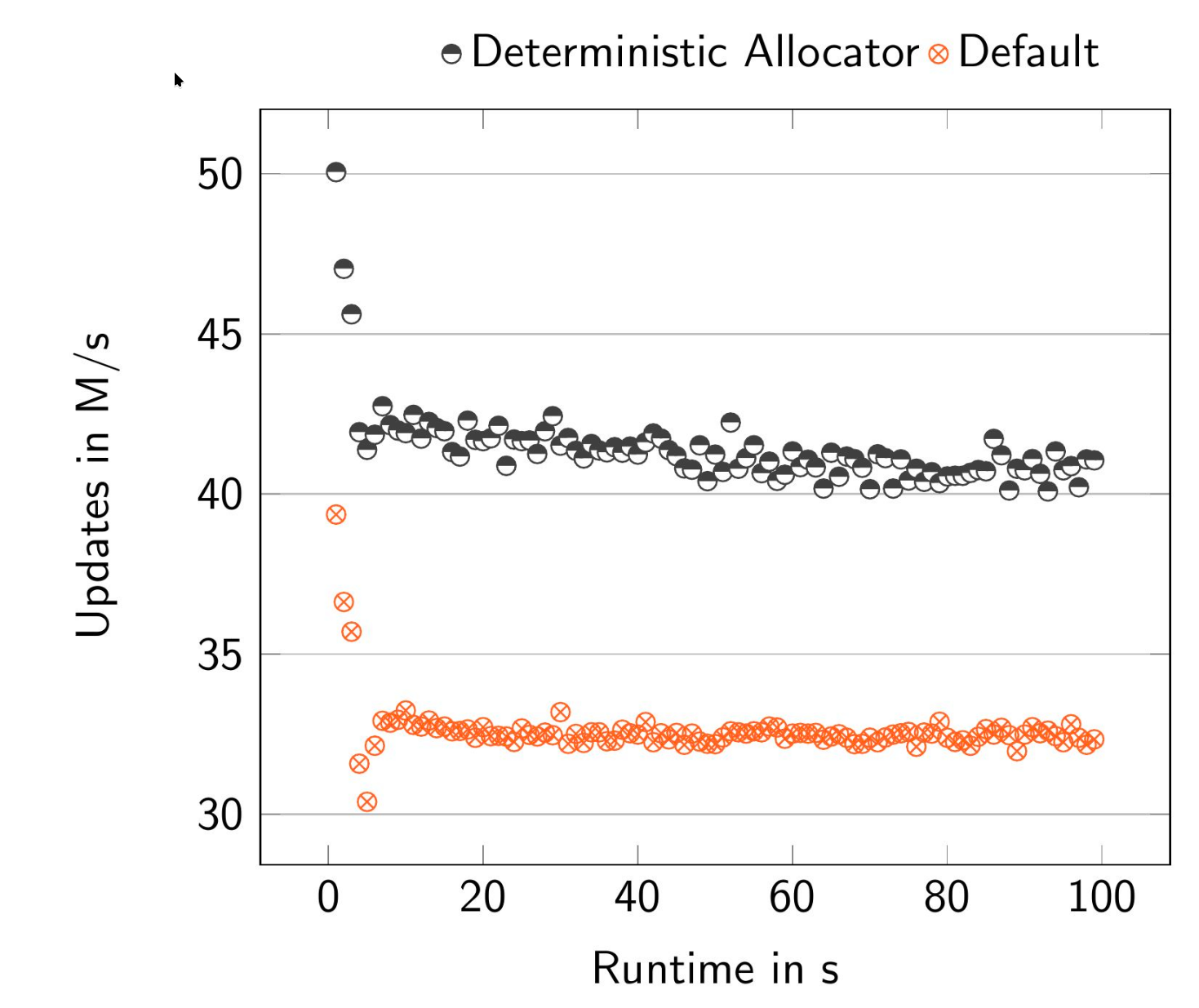}
    \caption{The effects of enabling the deterministic allocator for huge (2MiB) pages on AWS Intel Sapphire Rapids system.}
  \end{subfigure}
  \begin{subfigure}{0.3\linewidth}
    \includegraphics[width=\textwidth]{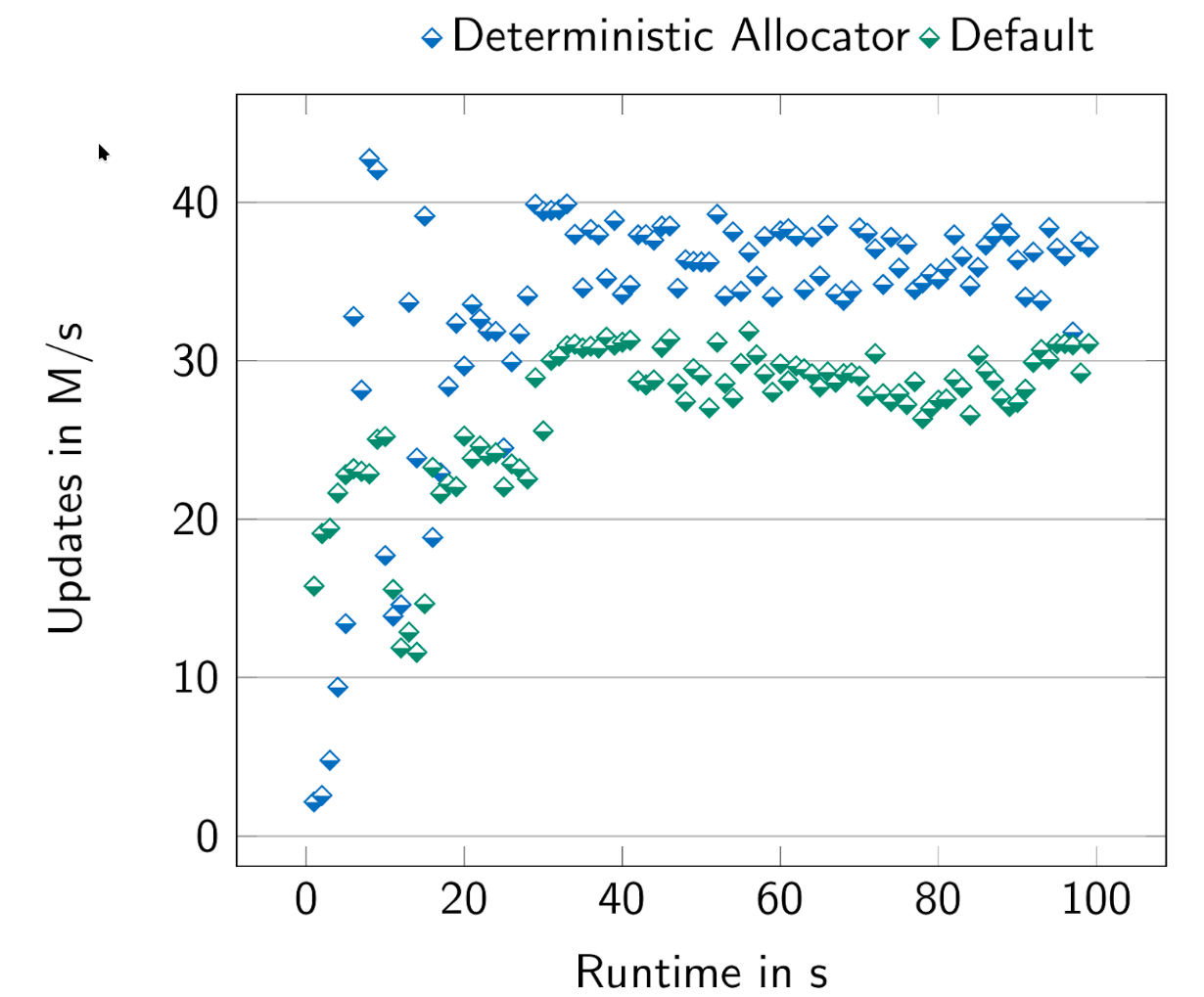}
    \caption{The effects of enabling the deterministic allocator and adding prefetch instructions for default (4KiB) pages on AWS Intel Sapphire Rapids system.}
  \end{subfigure}
  \begin{subfigure}{0.3\linewidth}
    \includegraphics[width=\textwidth]{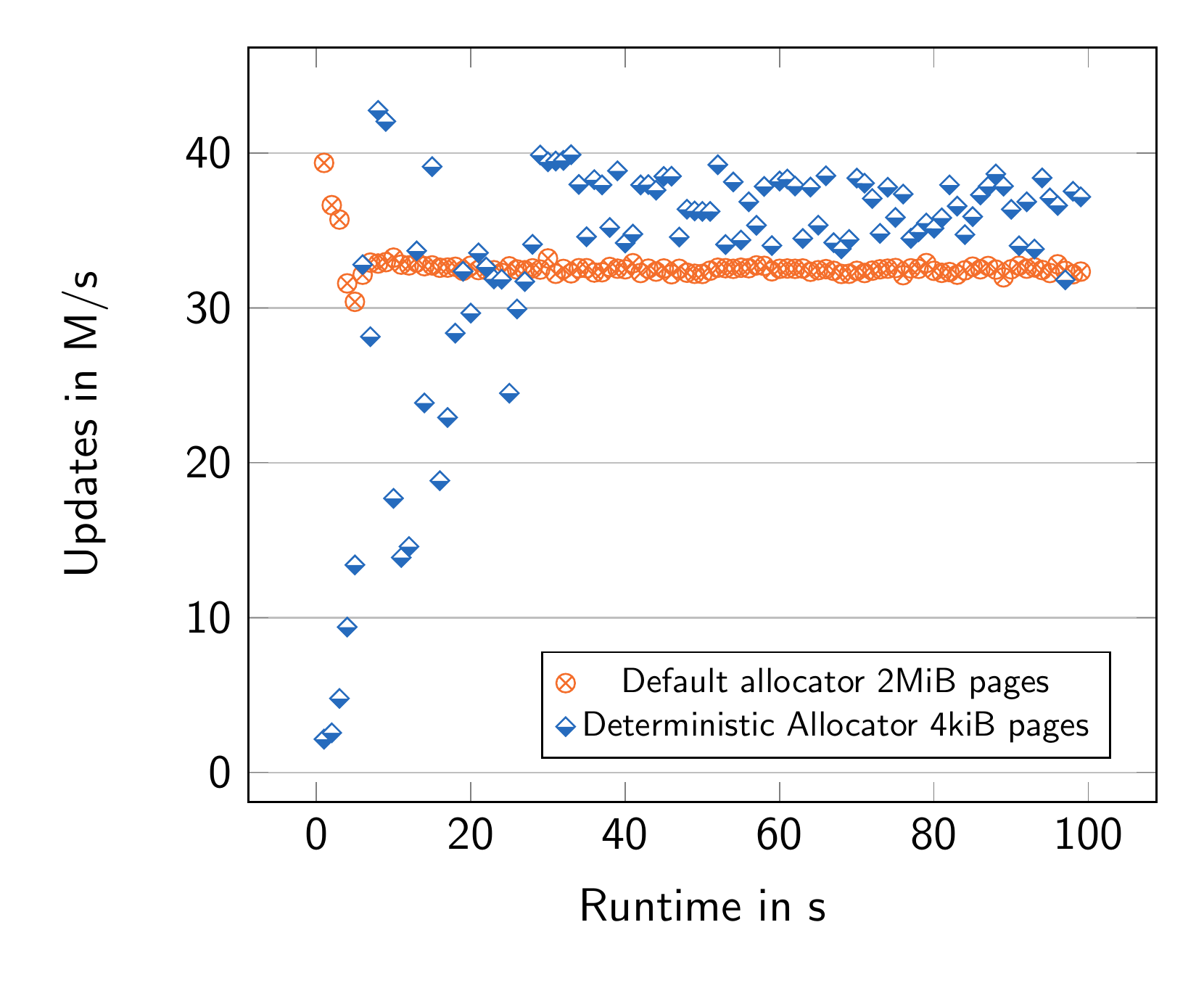}
    \caption{Comparison of relative effects of enabling huge pages and deterministic allocation (with appropriate prefetch instructions) on AWS Intel Sapphire Rapids system.}
  \end{subfigure}

  \caption{Evaluating prefetching.}%
  \label{fig:prefetching}
\end{figure*}

\subsubsection{Varying the Number of Sub-trees}
Inspired by other authenticated stores, \tool{} avoids re-hashing the Merkle tree all the way to the main root after every state update.  Instead, we compute hash changes in parallel, in per-thread  subtrees.  This means that we only compute the top levels of the tree once per block.  Since nodes closer to the root change more frequently, this dramatically reduces redundant hashing work.

Figure~\ref{fig:varying-subtrees} demonstrates this principle.  Increasing the number of subtree roots reduces the pressure on Pleiades, the in-memory component.  The optimal number of twig roots is determined by a few key factors.  The latency budget determines the upper bound on the amount of twig roots.

Thus finding the right number of sub-tree roots should be viewed as a multi-dimensional optimization problem.  It is determined by a combination of parameters, such as the snapshot frequency, the topology of the authenticated data store, the latency budget, and the ratio of CPU core count and single-threaded performance.  Consider that in our tests (Figure~\ref{fig:varying-subtrees}) we observe a notable 37\% difference in throughput based solely on adjusting the number of subtrees.

\subsubsection{Effects of Prefetching}
In its current iteration, \tool{} is a purely CPU-based authenticated data store.  Consequently, its performance is constrained by the inherent limitations of modern CPU architectures.  Specifically, when data is not readily available in the local cache, the cache is “thrashed”, \textit{i.e.}\ the CPU is not fed data optimally, and data loaded into the CPU caches has to be loaded from the comparatively slower DRAM\@.

Given these constraints, understanding the data processing timeline significantly boosts CPU efficiency.  Our central data structure, a Merkle Tree, is designed to pre-determine which nodes require traversal.  This foresight is crucial, considering the Pleiades' potential to house billions of nodes.  To this end, \tool{} employs a deterministic, breadth-first allocation order for nodes.  This is then followed up by issuing explicit prefetch instructions, which on some architectures (\textit{e.g.}  Zen~4) is disabled for security.

Figure~\ref{fig:prefetching} demonstrates the effects of enabling the prefetching (via deterministic allocation) on the overall performance of \tool.  The effect is most easily seen when the allocation takes place in~2MiB pages (Figure~\ref{fig:prefetching}a).  We observed a~25\% increase in throughput given the deterministic allocation.  The same effect can be observed on Figure~\ref{fig:prefetching}, where due to TLB walks, the data points are less concentrated, but the averages are still approximately~20\% apart.  It should be noted that enabling larger pages improves performance overall, but to a significantly lesser extent (as seen in Figure~\ref{fig:prefetching}).

It is also important to note that while optimizations like deterministic allocation for memory prefetching show significant gains (up to~25\%), the underlying prefetch instruction features may be disabled on common cloud platforms for security reasons.  For example, our main cloud provider OVH disabled this on our original AMD Zen~4 testbench.  AWS, thankfully, allows deterministic allocation to take place on our Intel Sapphire Rapids test machine, as that specific architecture is not vulnerable in the same way as Zen~4.  This achieves~48M updates per second.  Therefore, deploying \tool in a production environment requires a careful assessment of the available hardware and software stack to maximize performance.

\subsection{Experimental Summary}
Our experiments reveal that optimizing snapshot frequency, enabling prefetching, and carefully managing the number of twig roots are crucial for maximizing \tool's performance.

Tuning snapshot to~500ms gives~$2\times$ over baseline state updates per second; disabling yields another~$2\times$.  Proper subtree root depth spans a~37\% throughput window, constrained by snapshot frequency.  Enabling deterministic allocation adds~25\%,~2 MB pages add~5–-10\%.

\tool is designed to meet the needs of a very fast blockchain, but its emergent flexibility suggests its applicability in other contexts. Going forward we will explore\tool in contexts like light nodes and trusted RPCs.  We shall highlight deployment-time concerns such data-center based desegregated storage options.

% Local Variables:
% jinx-languages: "en_US"
% TeX-master: "main"
% End:

\section{Discussion}
\label{sec:discussion}

\subsection{Data Desegregation and Running at Scale}%
\label{sec:de-segregation}

As the scale of blockchain processing grows, in terms of both the transactions-per-second as well as they account key space, the desire to store data beyond a single machine develops as well.

Transactions after being verified and executed produce changes to the internal state of the blockchain.  These changes are propagated in the form of account bundles.  These are key-value pairs ordered such that it is possible to recover every intermediate state of the blockchain with transaction granularity.  These alone do not constitute state commitments, and must be further processed by Pleiades.

%As Figure~\ref{fig:dataflow} illustrates, 
The in-memory component segregates the data into the Merkle tree and journals.  The Merkle tree is updated up to a limited depth, as long as transactions are streaming.  With some configurable period, the tree is updated completely, and a snapshot is taken, which persists the journals and inclusion proofs.  The files are then forwarded for further processing; this could be distributing inclusion proofs to light clients, responding to an optimistic challenge, \emph{etc.}

As one can see in Figure~\ref{fig:arch}, there is no \emph{direct} communication between Hyades and Pleiades. This allows Pleiades to operate on a different machine or machines compared to Hyades.  This is useful for reducing disk contention, would allow the system to scale out, and potentially expand the overall system with archival machines as well.  

Because the snapshot files are versioned standard techniques for accelerating data storage such as striping are effective. The striping can be done at the software level, hardware level (RAID) or network level.  For instance, if the latency introduced by writing snapshots to disk is significant, one can have a central scheduler, \textit{e.g.} round robin to dispatch different snapshots to different networked machines such that the Pleiades component is never blocked. 

Furthermore, so-called ancient data~\cite{liangMoltDBAcceleratingBlockchain2024} does not demand the same level of granularity and scrutiny. One can dedicate separate machines to archiving ancient data, compressing it.  For instance, one can use the fact that ancient data does not necessarily need to be stored with the same granularity as current data, for example, by eliminating \(ABA\) patterns reducing them to just \(A\).  The journals being append-only also allow room for compression. 
\begin{comment}
\begin{figure}[tb]
  \centering
  \includegraphics[width=\linewidth]{figures/data-flow.png}
  \caption{Data flow in \tool.}%
  \label{fig:dataflow}
\end{figure}
\end{comment}

\subsection{Comparing with QMDB}
We used the publicly available version of QMDB from GitHub to do some of the measurements below.  We used the same AWS machine as the one described above; this machine is more powerful than the machines we saw used in the QMDB paper.  The QMDB paper reports update numbers that reach~2.28M state updates per second.  Sadly, despite our efforts, we were unable to replicate the same numbers using the code in the QMDB repository\footnote{\url{https://github.com/LayerZero-Labs/qmdb}} (we call this QMDB~v0.2.0), although there may be some configuration differences accounting for these discrepancies.  We also measured against a more recent code drop of QMDB\footnote{\url{https://github.com/LayerZero-Labs/fafo/tree/main/qmdb}} (from June~26,~2025) for completeness; we call this QMDB in the table below.

In our measurements we obtain~1,280,000~TPS for v1 and~803,000~TPS for v2.  We used the provided \texttt{speed} and \texttt{bench-sendeth} tools to generate workloads utilizing random keys.  We see improvement of at least an order of magnitude for \tool compared to QMDB. We hypothesize that some of the differences could be caused by not being able to scale beyond a fixed number of cores and excessive use of locking.
\section{Related Work}
\label{sec:related}
In this section, we go beyond the the projects that are most immediately relevant for \tool, such as LVMT~\cite{lvmt} and QMDB~\cite{qmdb}. 

\subsection{Acceleration Approaches for Merkle Trees}
El-Hindi~\etal~\cite{el-hindiMerkleTreesHighPerformance2023} demonstrate adapting Merkle trees for the parallel processing power of contemporary CPUs. Classical data structures such as B-Trees perform millions of operations per second, whereas Merkle Trees manage merely tens of thousands due to two key reasons. First, updates in Merkle Trees require CPU-heavy tasks, like computing cryptographic hashes or digital signatures, constraining performance to CPU speed. Second, enhancing performance through multithreading is difficult; as the root hash is updated each time, threads conflict at the root necessitating synchronization, leading to high contention and reduced performance on multi-core systems. The paper introduces an efficient synchronization method for Merkle Trees and proposes \emph{a reverse latch-coupling} to boost concurrency. Furthermore, a splitting concept is suggested to decrease both contention and cryptographic overhead simultaneously.

Wang~\etal~\cite{wangExampleParallelMerkle2024} propose a parallel Merkle tree traversal method focusing on optimizing a single task's efficiency for algorithmic parallelism, and achieving full GPU performance for data parallelism. Notably, they introduce the first CPU-GPU collaborative solution aimed at minimizing CPU-GPU communication overhead in traversal algorithms. Their results show our algorithm is still $4.48\times$ faster in algorithmic parallelism than the ideal benchmark. For data parallelism, raising core count from~1 to~8,192 yields a parallel efficiency of~78.39\%.

Deng~\etal~\cite{dengAcceleratingMerklePatricia2024} present a novel approach to accelerate MPT by leveraging GPU parallelism. The challenge involves (i) the complexity of lock-free data structures and (ii) the unscalability of traditional fine-grained locking on GPU. They investigate issues in GPU-based MPT acceleration, focusing on node splitting and hash computation conflicts during parallel insertions. The authors propose PhaseNU, a lock-free algorithm, and LockNU, a lock-based algorithm, to resolve node splitting conflicts, accompanied by a decision model to choose the appropriate algorithm for different workloads. Furthermore, they introduce PhaseHC, a GPU-based algorithm, to avoid hash computation conflicts. These techniques are validated by integrating them into Geth and LedgerDB and testing on two real and one synthetic dataset, achieving significantly better performance than existing MPT solutions in Geth, with~$19.79\times$ and~$29.69\times$ speedups for insertion.

Tian~\etal~\cite{tianLETUSLogStructuredEfficient2024} 
present LETUS, a Log-structured Efficient Trusted Universal Storage designed for blockchain applications, delivering cryptographic tamper-evidence along with high performance and resource efficiency. The authors highlight that 
(i) Tree encoding in the two-layered architecture (LSM-based key-value store plus an ADS) leads to severe I/O and space amplification. Each tree node is encoded as a key-value pair in the backend storage, resulting in the complete rewriting of the entire node even when only a single object is modified. (ii) The hash-based indexing schema leads to substantial disk bandwidth consumption. Nodes in the tree are addressed by the Merkle hash derived from the content of the node itself. The randomness of hash leads to a random distribution of keys, necessitating compactions of SST-files in LSM-tree-based storages to keep keys sorted. These background compactions significantly consume disk bandwidth. (iii) The two-layered architecture introduces a lengthy I/O path. Accessing the state of a single account requires multiple node queries in the ADS, triggering multiple I/O operations in the backend database to search for key-value pairs across the multi-level SST-files.
Specifically, (1) LETUS integrates the Authenticated Data Structure (ADS) directly into the storage engine, overcoming the conventional two-layered architecture to facilitate detailed I/O optimizations. (2) LETUS introduces the DMM-Tree, a new ADS that merges Merkle tree capabilities with delta-encoding to notably decrease storage requirements. (3) A version-based indexing system is employed by LETUS to handle the sizable volume of pages created by ADS, utilizing a B-tree variant for effective page storage indexing. (4) LETUS offers a universal solution applicable to varied blockchains, including public options like Ethereum, BNB Smart Chain, and examples of consortium blockchains like AntChain. It has been implemented in commercial applications of AntChain, such as NFTs. 
The experimental evaluation indicates that LETUS enables AntChain to boost throughput by up to~$15.8\times$ and cut storage costs by~80.3\%, while Ethereum sees up to~$10.1\times$ throughput enhancement and~75.0\% storage savings.

%\cite{amiriParBlockchainLeveragingTransaction2019, bappyMaximizingBlockchainPerformance2024, capocasaleParallelTransactionExecution2024, chenPEEPParallelExecution2021,guptaProofofExecutionReachingConsensus2021, huEfficientBlockchainDecentralized2025, shahidParallelTransactionExecution, wuFollowingThreadFinding2024,huEVMTracerDynamicAnalysis2023,schleiersmithReStreamAcceleratingBacktesting2016}

\subsection{High-integrity Storage}
In their work, Arasu~\etal~\cite{arasuFastVerMakingData2021} introduce FastVer, a performant key-value store ensuring robust data integrity. Built as an enhancement to the open-source FASTER, FastVer retains FASTER's key-value API and introduces a \texttt{verify()} function to identify unauthorized database alterations and ensure consistent read operations with historical updates. FastVer leverages a novel methodology combining Merkle trees with deferred memory verification, achieving throughput significantly higher than traditional Merkle tree or memory verification techniques. Formal proofs validate this method's correctness, ensuring \texttt{verify()} functions accurately, barring cryptographic hash collisions. Performance tests reveal FastVer handles over~50 million key-value operations per second, with swift verification for diverse databases and workloads, rivaling FASTER without data integrity and surpassing systems like Redis. FastVer outperforms Merkle tree-based systems by two magnitudes in performance and exceeds Concerto~\cite{arasuConcertoHighConcurrency2017} by one order of magnitude in both throughput and latency.

Burke~\etal~\cite{burkeScalableIntegrityChecking2025} 
observe that while Merkle hash trees are a standard for ensuring the integrity and freshness of stored data, they introduce significant compute and I/O costs, particularly on the I/O critical path—a challenge that has not been fully characterized by previous research. In this paper, the authors quantify these performance overheads in realistic settings to better understand their impact.
Their analysis of the root causes of these overheads led them to design a new, optimized tree structure called Dynamic Merkle Trees (DMTs).
Traditional hash trees are often balanced, but research on real-world workloads reveals a different reality: they are highly skewed, with a small number of data blocks being accessed much more frequently than others. This skewed access pattern means that an optimal hash tree is often far from balanced. Instead, frequently accessed blocks should have shorter verification and update paths within the tree. 
DMTs address this by adapting on the fly. Building on the splay trees used in garbage collection and IP routing, DMTs are a novel dynamic, unbalanced hash tree design that self-adjusts at runtime. This allows them to learn and adapt to workload patterns without prior knowledge, continuously reducing hashing costs for frequently accessed data.
The authors tested their approach using Zipfian workloads, an Alibaba dataset, and a Filebench OLTP workload. Their results show that the static nature of state-of-the-art approaches is a major bottleneck, delivering less than~50\% of optimal throughput on average.
In stark contrast, DMTs capitalize on skewed access patterns to deliver over~85\% of optimal throughput. This translates to a~$2.2\times$ improvement in throughput and latency over existing solutions. The findings  in the paper demonstrate that for cloud block storage, balanced trees are ill-suited. DMTs provide a promising new direction for achieving storage integrity efficiently and at scale.

Wang~\etal~\cite{wangV2FSVerifiableVirtual2024} highlight the critical need for sophisticated query techniques, such as data aggregation and correlation analysis, to derive valuable insights. However, executing queries across multiple blockchains presents challenges, including issues of blockchain compatibility, diverse query support, and maintaining query accuracy. The authors propose a novel solution called the verifiable virtual filesystem (V2FS) to overcome these obstacles. V2FS extends the conventional POSIX I/O interface and transitions the focus from computational verification to data verification. This approach allows clients to employ standard database engines to perform queries with verifiable data provided by an ISP. The solution ensures robust data integrity and integrates effortlessly with existing systems to support a variety of query types. To address blockchain compatibility, the DCert framework is employed to certify blocks from different blockchains, thereby enhancing the system's applicability. Additionally, they introduce algorithms based on caching and bloom filters to improve query performance and reduce network expenses. Comprehensive security analysis and empirical evaluations confirm the effectiveness and efficiency of the system.

Yue~\etal~\cite{yueGlassDBEfficientVerifiable2023} delve into the architectural space of verifiable ledger databases by examining three primary dimensions: abstraction, threat model, and performance capabilities. Their examination of existing systems reveals two primary drawbacks: the absence of transactional functionality and subpar efficiency. To address these challenges, the authors introduce GlassDB, an innovative distributed database system that mitigates these gaps within a realistic threat model framework. GlassDB maintains the core feature of verifiability seen in transparency logs, while simultaneously facilitating transactional support and enhancing performance levels. It achieves this by augmenting a ledger-like key-value store with a proficient data structure for proof generation and incorporates a concurrency control mechanism to enable transactions. When updating its fundamental data structures, GlassDB efficiently processes independent operations arising from concurrent transactions by batching them. Furthermore, the authors expand the benchmarking landscape for verifiable ledger databases by integrating the YCSB and TPC-C benchmarks, thereby creating a novel evaluation standard. This enhanced benchmark is utilized to perform comparative analysis, pitting GlassDB against four baseline systems: three re-implemented versions of existing verifiable databases and a verification map grounded in a transparency log. The experimental findings underscore that GlassDB stands out as an effective, transactional, and verifiable ledger database system.
% Local Variables:
% jinx-languages: "en_US"
% TeX-master: "main"
% End:

\section{Conclusions}%
\label{sec:conclusions}

\tool signifies a significant advancement in the domain of Authenticated Database Design. By eliminating disk I/O from the critical execution path, implementing prefetching, and optimizing the update pattern of the Merkle tree, we have developed an ADS capable of processing state updates at a 50 Gbps network line rate. This throughput substantially exceeds any practically observed blockchain throughput, ensuring that even the most rapid, high-throughput blockchains can generate state commitments. 

Additionally, \tool facilitates historical state proofs, thereby supporting a multitude of new applications. For instance, the use of \tool could enable blockchains lacking state commitments to support light clients and/or rollups. Further applications include the facilitation of Time-Weighted Average Price aggregation at the blockchain's tip, with any arbitrary time granularity. Moreover, \tool exhibits superior performance and simplicity compared to other ADS projects. 

Specifically, \tool is at least an order of magnitude faster than QMDB, achieving 48M updates per second on the same hardware with similar configuration settings. Notably, \tool relies on just three external packages, while QMDB necessitates~21 external dependencies. 
This makes \tool a viable option for environments with extremely constrained resources, particularly since its historical data functionality can be isolated and turned off, further enhancing performance. This capability has been demonstrated on consumer-grade portable hardware, achieving~8M updates per second in a purely in-memory mode and~5M updates per second with sub-second snapshot periodicity, thereby illustrating attractive and cost-effective scalability.

\point{Acknowledgments} The authors are most grateful to {Olivier Desenfans}, and {Nagaraju Thogiti} for their most valuable input during the writing of this paper.

%\newpage
\bibliographystyle{plain}
\bibliography{ref}

\begin{thebibliography}{10}

\bibitem{blockstm}
Fikunmi Ajayi-Peters.
\newblock {B}lock-{S}{T}{M} vs. {S}ealevel: {A} {C}omparison of {P}arallel {E}xecution {E}ngines --- eclipse.xyz.
\newblock \url{https://www.eclipse.xyz/articles/block-stm-vs-sealevel-a-comparison-of-parallel-execution-engines}, 2024.
\newblock [Accessed 07-07-2025].

\bibitem{arasuFastVerMakingData2021}
Arvind Arasu, Badrish Chandramouli, Johannes Gehrke, Esha Ghosh, Donald Kossmann, Jonathan Protzenko, Ravi Ramamurthy, Tahina Ramananandro, Aseem Rastogi, Srinath Setty, Nikhil Swamy, Alexander Van~Renen, and Min Xu.
\newblock {{FastVer}}: {{Making Data Integrity}} a {{Commodity}}.
\newblock In {\em Proceedings of the {{International Conference}} on {{Management}} of {{Data}}}, 2021.

\bibitem{arasuConcertoHighConcurrency2017}
Arvind Arasu, Ken Eguro, Raghav Kaushik, Donald Kossmann, Pingfan Meng, Vineet Pandey, and Ravi Ramamurthy.
\newblock Concerto: {{A High Concurrency Key-Value Store}} with {{Integrity}}.
\newblock In {\em Proceedings of the {{ACM International Conference}} on {{Management}} of {{Data}}}, 2017.

\bibitem{blake2}
{Jean Philippe} Aumasson, Willi Meier, {Raphael C.W.} Phan, and Luca Henzen.
\newblock {\em BLAKE2}, pages 165--183.
\newblock Information Security and Cryptography. Springer, Switzerland, 2014.

\bibitem{rise}
Samuel Battenally, Hai Nguyen, and Thanh Nguyen.
\newblock {RISE}: The gigagas layer 2.
\newblock {\em RISE labs}, May 2024.

\bibitem{burkeScalableIntegrityChecking2025}
Quinn Burke, Ryan Sheatsley, Rachel King, Owen Hines, Michael Swift, and Patrick McDaniel.
\newblock On {{Scalable Integrity Checking}} for {{Secure Cloud Disks}}.
\newblock 2025.

\bibitem{dengAcceleratingMerklePatricia2024}
Yangshen Deng, Muxi Yan, and Bo~Tang.
\newblock Accelerating {{Merkle Patricia Trie}} with {{GPU}}.
\newblock 17(8), 2024.

\bibitem{el-hindiMerkleTreesHighPerformance2023}
Muhammad El-Hindi, Tobias Ziegler, and Carsten Binnig.
\newblock Towards {{Merkle Trees}} for {{High-Performance Data Systems}}.
\newblock In {\em Proceedings of the {{Workshop}} on {{Verifiable Database Systems}}}, 2023.

\bibitem{gelashviliBlockSTMScalingBlockchain2022}
Rati Gelashvili, Alexander Spiegelman, Zhuolun Xiang, George Danezis, Zekun Li, Dahlia Malkhi, Yu~Xia, and Runtian Zhou.
\newblock Block-{{STM}}: {{Scaling Blockchain Execution}} by {{Turning Ordering Curse}} to a {{Performance Blessing}}.

\bibitem{lvmt}
Chenxing Li, Sidi~Mohamed Beillahi, Guang Yang, Ming Wu, Wei Xu, and Fan Long.
\newblock {LVMT}: {An} {Efficient} {Authenticated} {Storage} for {Blockchain}.
\newblock {\em ACM Transactions on Storage}, 20(3), August 2024.

\bibitem{liangMoltDBAcceleratingBlockchain2024}
Junyuan Liang, Wuhui Chen, Zicong Hong, Haogang Zhu, Wangjie Qiu, and Zibin Zheng.
\newblock {{MoltDB}}: {{Accelerating Blockchain}} via {{Ancient State Segregation}}.
\newblock 35(12):2545--2558.

\bibitem{tianLETUSLogStructuredEfficient2024}
Shikun Tian, Zhonghao Lu, Haizhen Zhuo, Xiaojing Tang, Peiyi Hong, Shenglong Chen, Dayi Yang, Ying Yan, Zhiyong Jiang, Hui Zhang, and Guofei Jiang.
\newblock {{LETUS}}: {{A Log-Structured Efficient Trusted Universal BlockChain Storage}}.
\newblock In {\em Companion of the {{International Conference}} on {{Management}} of {{Data}}}, 2024.

\bibitem{wangV2FSVerifiableVirtual2024}
Haixin Wang, Cheng Xu, Xiaojie Chen, Ce~Zhang, Haibo Hu, Shikun Tian, Ying Yan, and Jianliang Xu.
\newblock {{V2FS}} : {{A Verifiable Virtual Filesystem}} for {{Multi-Chain Query Authentication}}.
\newblock In {\em {{International Conference}} on {{Data Engineering}} ({{ICDE}})}. Ieee, 2024.

\bibitem{wangExampleParallelMerkle2024}
Ziheng Wang, Xiaoshe Dong, Yan Kang, Heng Chen, and Qiang Wang.
\newblock An {{Example}} of {{Parallel Merkle Tree Traversal}}: {{Post-Quantum Leighton-Micali Signature}} on the {{GPU}}.
\newblock 21(3), 2024.

\bibitem{solana}
Anatoly Yakovenko.
\newblock Solana: {A} new architecture for a high performance blockchain v0.8.13.
\newblock 2018.

\bibitem{yueGlassDBEfficientVerifiable2023}
Cong Yue, Tien Tuan~Anh Dinh, Zhongle Xie, Meihui Zhang, Gang Chen, Beng~Chin Ooi, and Xiaokui Xiao.
\newblock {{GlassDB}}: {{An Efficient Verifiable Ledger Database System Through Transparency}}.
\newblock 16(6), 2023.

\bibitem{qmdb}
Isaac Zhang, Ryan Zarick, Daniel Wong, Thomas Kim, Bryan Pellegrino, Mignon Li, and Kelvin Wong.
\newblock {QMDB}: {Quick} {Merkle} {Database}.
\newblock {\em Layer Zero Labs}, January 2025.
\newblock arXiv:2501.05262 [cs].

\end{thebibliography}

\end{document}